# Impedance spectroscopy: Impedance spectroscopy of nanomaterials

Rainer Schmidt[1,2,x]

[1] *Universidad Complutense de Madrid, GFMC, Dpto. Física Aplicada III, Facultad de Ciencias Físicas, 28040 Madrid, Spain*

[2] *Unidad Asociada "Laboratorio de heteroestructuras con aplicación en espintrónica", UCM/CSIC, Sor Juana Inés de la Cruz, 3, Cantoblanco E-28049 Madrid, Spain*

[x] email address: rainerxschmidt@googlemail.com

**Abstract**

Solid state impedance spectroscopy enables the various contributions to the resistive and capacitive properties of electronically inhomogeneous condensed matter to be deconvoluted and characterized separately. The different contributions arise from electronically distinct areas in the sample, which can in an ideal case be represented each by one standard RC element. In the following, the basic principles of impedance spectroscopy, different types of experimental setups and several examples of experimental impedance data sets from nanostructured materials are reviewed and discussed. The data analysis and equivalent circuit modelling processes which are relevant for the application of this technique to nanomaterials are emphasized. The different dimensions and structure of nanomaterials as compared to macroscopic bulk samples leads to quite different and sometimes more complex data that require detailed analysis and advanced equivalent circuit models.

# Impedance Spectroscopy

## *Impedance Spectroscopy of Nanomaterials*

*Rainer Schmidt*

## Contents

### I. Introduction

The scope of impedance spectroscopy (IS) measurements has been extended to broader areas of research continuously during the last decades. Originally, the method was mainly used to characterise liquid phase electrochemical cells, but its power to characterise resistive and capacitive properties of solid matter has been exploited more frequently ever since.[1] The analysis and equivalent circuit modelling of impedance spectroscopy data obtained from macroscopic bulk materials is performed routinely and is well understood at these days.[2, 3] On the other hand, the application of impedance spectroscopy to nanomaterials is becoming increasingly popular only recently, where the main aspiration is to achieve local impedance characterization on small length scales down to the nano-scale.

Macroscopic bulk materials are most conveniently characterized by IS in a parallel-plate capacitor measurement geometry, where the measured resistance $R$ and capacitance $C$ allow a quick and easy characterization of the resistive and capacitive properties respectively of different areas in the sample. The specific parameters of resistivity $\rho$ and dielectric permittivity $\varepsilon_r$ can be calculated readily by the well-known Equations:

$$\rho = R\frac{A}{d} = R\ g; \quad \varepsilon_r = \frac{C}{\varepsilon_0}\frac{d}{A} = \frac{C}{\varepsilon_0 g}; \tag{1}$$

where $A$ is the effective current cross section, $d$ the electrode distance, $g$ is the geometrical factor ($g = A/d$) and $\varepsilon_0$ the dielectric permittivity of vacuum. In the parallel-plate capacitor geometry, $A$ corresponds to the electrode area. However, this geometry sometimes can or cannot be achieved for nanomaterials with patterned, thin film or molecular structure. In the latter case, further considerations and possibly new models may be required to adequately determine and interpret their dielectric properties as measured by IS. In both cases, the geometrical factor varies significantly as compared to macroscopic bulk, and concomitantly the impedance spectra differ.

In the first section of this paragraph, the basic principles of IS are briefly reviewed, with special emphasis on the parts that are different for the application of IS to nano-structured materials. In particular, the rather simplified brick-work layer model applicable to polycrystalline bulk materials often requires modification for nano-structured materials, where different contributions are not always connected in series as is the case for macroscopic bulk samples. In the second section, several measurement geometries with different *g* values for different types of thin film structures are discussed. The third section is dedicated to several experimental examples on IS applied to nanomaterials, including data analysis and equivalent circuit modelling.

## II. Basic principles of impedance spectroscopy

### a) Impedance response on the phasor diagram

There are two ways to perform impedance spectroscopy (IS): In the time ($t$) dependent domain a step function of voltage is applied to a measurement cell or electrode - sample configuration. As a consequence, the electric dipoles in an insulating material polarise. After the voltage signal is switched off abruptly, the time dependent current signal due to the depolarisation or relaxation of the sample is measured and such time-dependent results are usually Fourier transformed into the frequency domain. Therefore, it is more convenient to measure the impedance in the frequency domain directly, which does not involve major experimental difficulty nowadays by the use of state of the art broad frequency range impedance analysers. Furthermore, leakage currents in less insulating samples can be accounted for much easier in the frequency domain. The standard frequency domain measurement consists effectively of an electric stimulus in terms of a time ($t$)-dependent alternating voltage signal $U$ of sinus or cosine shape, with angular frequency $\omega$ and amplitude $U_0$ applied to the sample, and effectively the amplitude $I_0$ and phase shift $\delta$ of the current response signal $I$ are measured.

$$U(\omega, t) = \mathbf{U_0} \cos(\boldsymbol{\omega} t); \rightarrow I(\omega, t) = I_0 \cos(\boldsymbol{\omega} t - \delta); \quad (2)$$

All parameters defined by the applied voltage signal ($U_0$, $\omega$) are printed in blue/bold, the measured parameters of the current response are in red ($I_0$, $\delta$). In Equation (2) it is assumed that the dielectric response is linear, i.e. the applied voltage and the current response signals have both the same shape and frequency. In the case of non-linear charge transport (i.e. in diodes or Schottky interfaces) or for non-linear dielectric permittivity (i.e. in ferroelectrics), the shape of the response signal may vary. However, most state-of-the-art impedance analysers can scope with such non-linearity to still reliably determine the phase shift and amplitude of the current response signal. Advanced impedance analysers allow more detailed analysis of the non-linear effects by displaying higher harmonics of the current response signal.

On the phasor diagram in Figure 1, one phase of the applied voltage signal corresponds to a $2\pi$ (360°) rotation of the voltage arrow $U(\omega, t)$. The current response of ideal circuit components is: (1) in-phase with the applied voltage in the case of an ideal resistor $R$; (2) out-of-phase by $\delta = -\pi/2$ for an ideal capacitor $C$; and (3) out-of-phase by $\delta = +\pi/2$ for an ideal inductor.

All phase angles depicted in Figure 1 are time independent and all arrows rotate at constant angles for a given frequency. The impedance can therefore be defined as a time-independent complex number $Z^*$ (= $Z'$ + i$Z''$), where the impedance fraction in-phase with $U(\omega,t)$ is defined as the real part $Z'$, the fraction +/- $\pi/2$ out-of-phase as the positive/negative imaginary part $Z''$, and $i = \sqrt{-1}$. The complex impedance $Z^*$ definitions for several electronic circuit components are given in Figure 1. Figure 1 also depicts the phase angle $\delta$ of a typical RC element consisting of a resistor and capacitor connected in parallel. This RC component constitutes the basis of equivalent circuits in bulk materials, but may be modified in nanomaterials as discussed below. It should be noted that an alternative nomenclature of the impedance $R$ (= $Z'$) and reactance $X$ (= $Z''$) are sometimes used in the literature ($Z^* = R +$ i$X$). The relationships of |Z| (modulus of the impedance), $Z'$, $Z''$, $U_0$, $I_0$ and the phase angle $\delta$ can be expressed as:

$$|Z| = \frac{U_0}{I_0} = \left(Z'^2 + Z''^2\right)^{1/2}; \quad Z' = |Z|\cos\delta; \quad Z'' = |Z|\sin\delta; \qquad (3)$$

### b) Non-ideal impedance response

In reality, the different dielectric contributions from different areas in a solid state sample are hardly ever ideal and, therefore, experimental IS data can usually not be modelled by using ideal RC elements. To account for the non-ideality of dielectric response, the ideal capacitors in the equivalent circuit are often replaced by constant-phase elements (CPEs). The phase angle and the complex impedance $Z_{CPE}^*$ of a CPE are shown in Figure 1, where $C_{CPE}$ is the CPE specific capacitance. The CPE capacitance is given in modified units of [Farad·$s^{n-1}$] and can be converted into a real capacitance given in [Farad] using a standard conversion.[4]

The critical exponent $n$ defined in Figure 1 has typical values of $n = 0.6 - 1$. $n = 1$ constitutes the ideal case of an ideal capacitor for an ideal dielectric relaxation. In a non-ideal equivalent circuit containing a CPE, decreasing $n$ values indicate a broadening of the respective dielectric relaxation peak as a reflection of the broadening of the distribution of dielectric relaxation times or time constants, $\tau$, across the sample.[5-8] In an ideal RC element $\tau$ is given by $\tau = R{\cdot}C$. The exact shape of the distribution of $\tau$ is complicated to be determined from IS data,[9] and the exponent $n$ constitutes a semi-empirical parameter to reflect an increasing width of the distribution in $\tau$ by decreasing $n$ values.

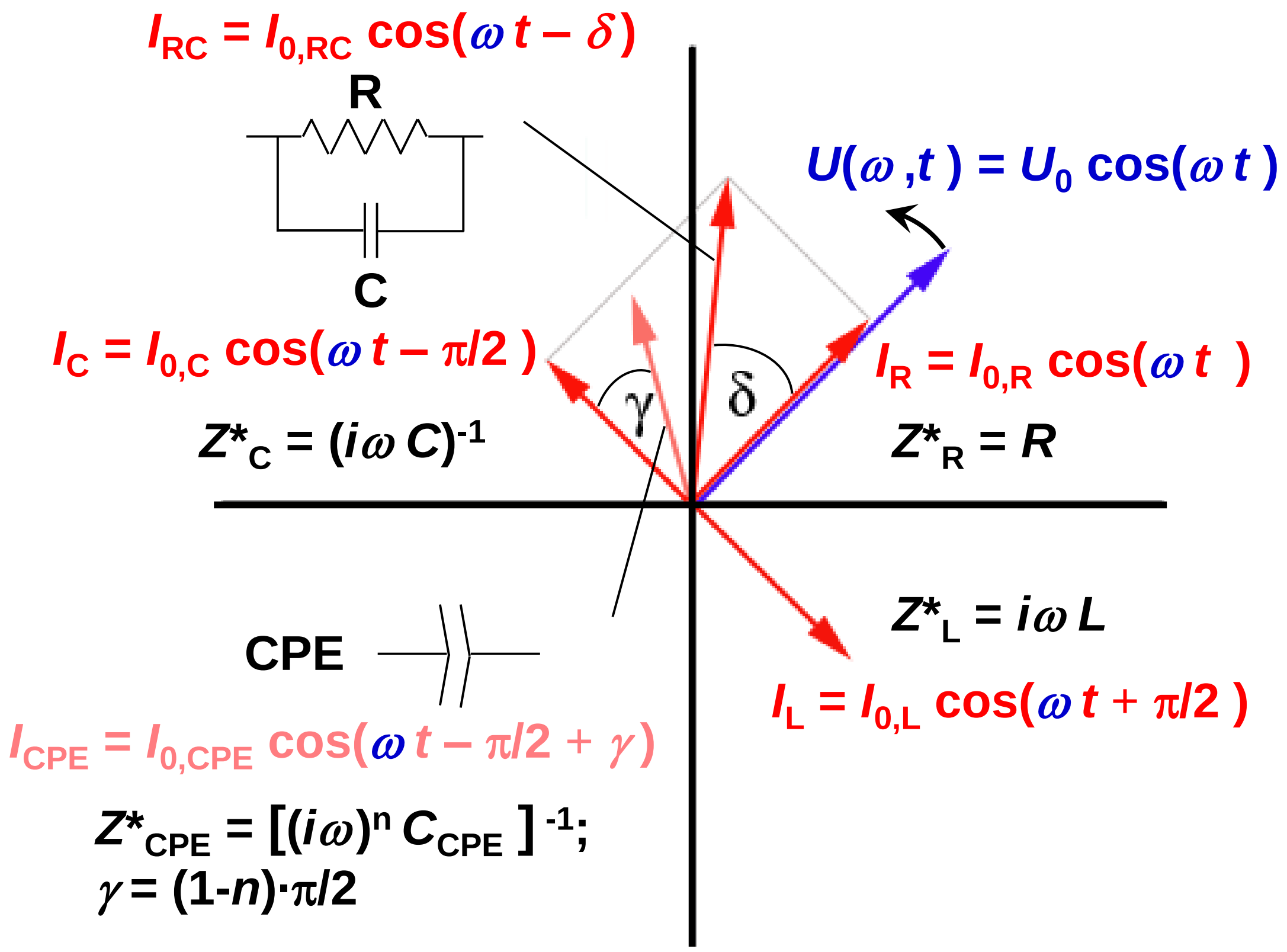

**Figure 1.** Impedance response of standard circuit components on a phasor diagram: applied voltage stimulus $U(\omega,t)$, current response $I_R$ from an ideal resistor, $I_C$ from an ideal capacitor, $I_{RC}$ from an ideal parallel resistor-capacitor (RC) element with phase shift $\delta$, $I_L$ from an ideal inductor, and $I_{CPE}$ from a constant phase element (CPE) for a non-ideal capacitor with a frequency independent phase shift $\gamma$ with respect to $I_C$.

**c) The RC element model**

As mentioned above, one ideal parallel resistor – capacitor RC element is well established to describe the impedance of an ideal dielectric contribution from one region within the sample, where the electronic behavior is equivalent all across such one region. This model works particularly well for insulators, where the capacitor describes the ability of a material to store charge and the parallel resistor describes the leakage current due to some un-trapped charge carriers bypassing the charge storage element. The charge carriers can be of any kind and the RC element model works for all standard electronic and ionic conduction mechanisms. One RC element thus describes the intrinsic resistance and capacitance of most solid matter and in the case of a parallel plate capacitor measurement configuration the intrinsic resistivity $\rho$ and dielectric permittivity $\varepsilon_r$ can be calculated straight forward from the sample geometry according to Equation (1). In polycrystalline bulk materials additional extrinsic dielectric contributions from electronically distinct areas in the sample commonly arise from grain boundary (GB) and electrode – sample interface (IF) regions, where at least the former can usually be represented by an additional RC element in series.[10-12] The macroscopic impedance is then represented by just the sum of the individual RC impedances, i.e. all RC elements can be regarded to be connected in series as illustrated in Figure 2. The bulk, GB and interface capacitance have typical values in macroscopic samples (approximate dimensions of 1 cm x 1 cm x 1 cm) as indicated in Figure 2, which allows their identification via the capacitance.[2] In nanomaterials this classification is only applicable if a parallel plate capacitor measurement setup is used and the measured capacitance is related to the sample geometry, i.e. the resistivity and specific capacitance or dielectric permittivity can be calculated. This may not always be possible and for different geometrical arrangements or orientations of electronically distinct regions or phases, different equivalent circuits may be applicable. The brick-work layer model in macroscopic and polycrystalline samples (Figure 2) is based on the following assumptions:[3, 13, 14]

1. All intrinsic grain interior bulk areas all across the sample must possess similar dielectric properties, in order to be represented by one single RC element for the entire sample. The same must be true for the extrinsic contributions from GB and electrode interface IF regions.
2. Intrinsic grain interior bulk areas must exhibit distinctively different dielectric behaviour as compared to the other regions in the sample (i.e. GB and electrodes), to allow dielectric discrimination. In the same way, GB and electrode interface regions must be dielectrically distinct to discriminate them. In electronically homogeneous samples, only one intrinsic bulk related contribution (1 RC element) will be visible. The key parameter here is the time constant $\tau$, i.e. the product of the resistance and the capacitance in the first RC element representing the intrinsic sample regions needs to be distinctively different to the time constants of the other sample regions to allow discrimination. This implies that different resistance values are sufficient to discriminate two sample regions with equal capacitance and vice versa.

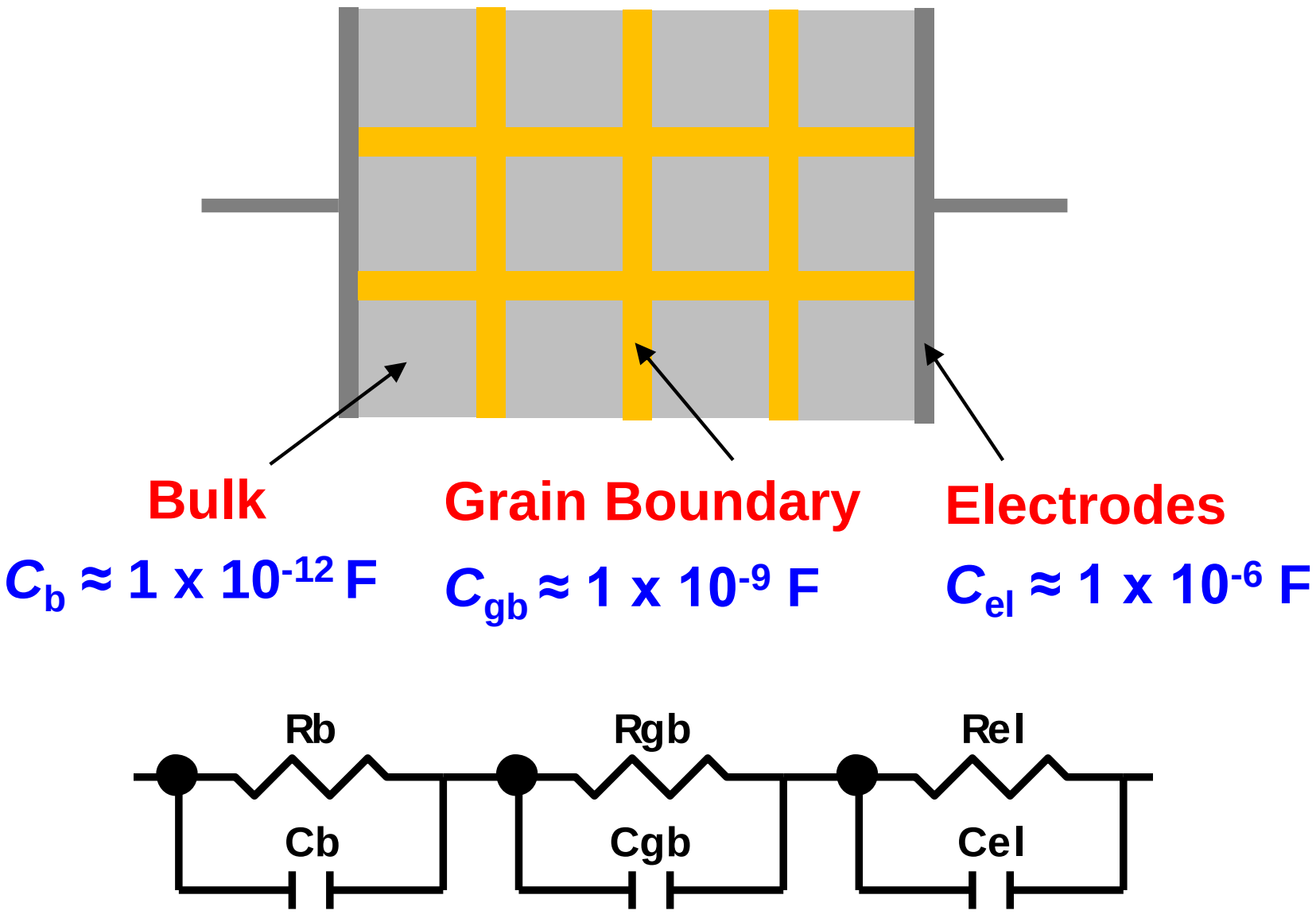

**Figure 2.** Brick work layer model for polycrystalline bulk materials.

3. All different regions in the sample are connected in series to each other. As shown in Figure 2, this may not always be precisely true in particular for the GB areas, which are in series and in parallel to grain interior bulk areas.
4. For a meaningful analysis of the resistivity of a sample, grain boundary areas should exhibit higher resistivity as compared to the grain interior bulk. Otherwise, charge carriers will propagate across the sample predominantly on pathways along the GBs, and neither the resistivity of the bulk nor of the GB areas can be calculated due to the ill-defined current cross section. In almost all polycrystalline materials this is not a problem and GB interfaces exhibit higher or at least equal resistivity and only few exceptions have been reported in the literature mainly on artificial interfaces.[15] This fact is relevant for criteria (c), because charge carriers can be expected to move across but not along GBs and the assumption of a series connection of bulk and GB areas may in fact be justifiable in most cases, i.e. a series connection of the respective RC elements usually represents a good model approximation.

In the case of nanomaterials it is clearly preferential to choose a parallel plate capacitor measurement configuration if possible, in which case the series RC element model may be valid. In any other case, the choice of the correct equivalent circuit may be less straight forward. It should be noted that different regions in a nanomaterial would still be represented by one RC element each, but possibly such RC elements needed to be connected to each other in different ways as discussed below. GB areas rarely exist in nanomaterials, e.g. in thin films, single molecules or other nano-sized objects, but electrode – sample interface contributions may in fact be quite common, e.g. in thin film structures, one film and one electrode contribution are commonly encountered in the IS data.[16-23]

**d) Experimental set-up for impedance spectroscopy (IS) measurements**

In the simplest case, a solid state IS experiment consists of a sample with two electrodes connected to two cables leading to the impedance analyser. This type of experiment is illustrated schematically in Figure 3, where an (empty) cubic shaped sample has two opposing faces of the cube covered with electrode material, represented by the black areas (as an electrode material usually noble metals are evaporated or sputter deposited to ensure optimal contact). The two electrodes are connected via two cables to the ports of an impedance analyser, in this case the 4192A Hewlett Packard Impedance Analyzer. The sample may be placed in a cryostat or/and in a magnetic field (indicated by the large dotted square around the sample) for temperature dependent or/and magnetic field dependent measurements. Figure 3 also illustrates that parasitic contributions can affect the impedance data obtained from the sample.

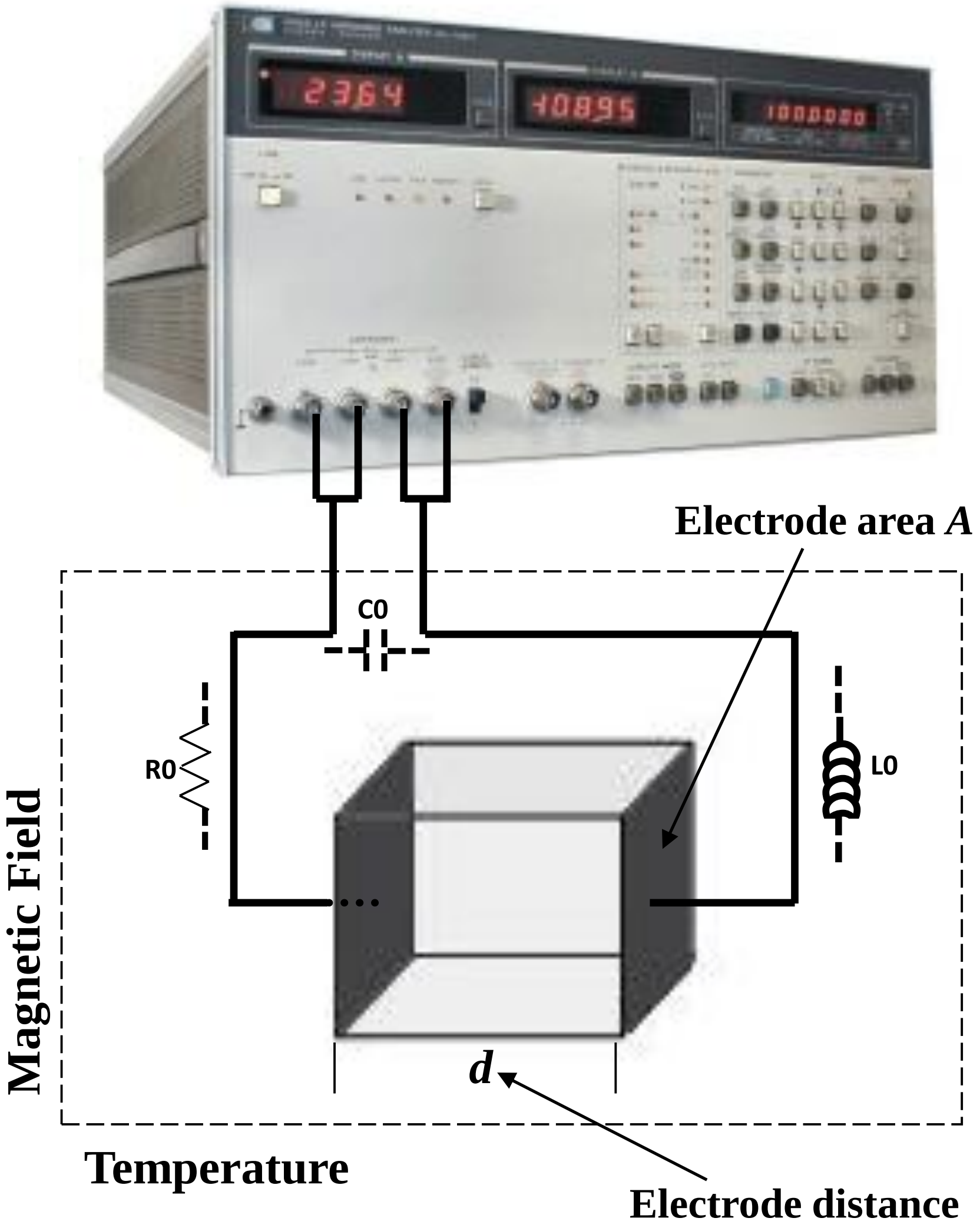

**Figure 3.** Impedance spectroscopy measurement set-up for a cubic shaped empty measurement cell representing a cubic sample. Possible parasitic contributions C0, R0 and L0 to the measured impedance are indicated.

In all IS measurements a parasitic capacitance C0 will occur as a result of the capacitance between the measurement cables and of the analyser. The two metallic measurement cables depicted in Figure 3 running to the impedance analyser with air in between them constitute a capacitor. In many measurement set-ups the measurement cables may in fact touch each other and are thus sheathed with insulating material. Again, this constitutes a parasitic capacitance, which can be minimised effectively by using shielded coaxial cables and C0 should not exceed the range of 0.2 – 0.8 pF. The measured capacitance obtained from the impedance analyser should be corrected for C0, which is connected in parallel to the true sample capacitance. C0 is not visible as a separate contribution but may affects the sample capacitance, and C0 should therefore be determined by previous open-circuit measurements. In the case that the sample capacitance is considerably larger than C0, the latter may be neglected. This is in fact the case for many nanomaterials of small dimensions.

Further parasitic contributions may arise from the resistance and inductance of the cables.[24] All measurement cables will always exhibit a certain parasitic resistance R0 and inductance L0, where the latter will necessarily dominate impedance spectra at high enough frequencies. Therefore, impedance spectroscopy data are usually reliable up to ≈ 1 MHz and usually unreliable above ≈ 10 MHz. The cable resistance is usually small as compared to the sample resistance of macroscopic bulk materials, but can be considerable in nanomaterials with small nominal sample resistances. Both parasitic contributions R0 and L0 may also contain contributions from the electrodes and are connected in series with the sample. To minimise the effects of L0 and R0, the co-axial measurement cables should be as short as possible. There are two possibilities to eliminate the parasitic components R0, L0 and C0 from the data analysis:

(a) The capacitance C0 can be measured by performing an open-circuit calibration measurement in advance, and the measured impedance from the sample in place can be corrected directly upon measurement by the impedance analyser or by the measurement software. In the same way, a short circuit calibration run (e.g. with touching measurement probes) allows eliminating R0 and L0.

(b) Parasitic contributions may not be eliminated from the data directly upon measurement, but can be accounted for by introducing the respective circuit components into the equivalent circuit model as demonstrated below.

The second "Uncorrected Impedance Response" method may be favourable, because it gives a quantitative measure on whether the experimental set-up is adequate. High R0, L0 and C0 values may indicate the necessity to improve the measurement set-up. Especially a high contribution from C0 can be critical, because it can dominate the spectra and mask sample capacitances leading to data misinterpretation. In the case that R0 and C0 may be negligible as compared to the sample resistance and capacitance, they can be removed from the equivalent circuit model. L0 would appear at high frequency and can often be effectively minimised or eliminated by restricting the measurement frequency range to values below 1 MHz.

Figure 3 demonstrates the preferable parallel-plate capacitor measurement configuration in terms of the effective current cross section $A$, which corresponds to the electrode area here, and electrode distance $d$, which allows simple calculation of the resistivity and dielectric permittivity according to Equation 1. For measurements in nanomaterials the effective current cross section $A$ can often not be determined reliably, in which case only nominal parameters such as resistance and capacitance can be analysed.

**e) Representations of IS data**

As mentioned above, an IS experiment consists of the measurement of the amplitude and phase shift of the current response of the sample to a voltage stimulus. From such measured parameters it is advantageous to obtain the real and imaginary parts of the impedance $Z'$-$Z''$, capacitance ($C'$-$C''$) or dielectric permittivity ($\varepsilon'$-$\varepsilon''$), the complex conductivity ($\sigma'$-$\sigma''$), and the imaginary part of the dielectric modulus ($M''$). Plotting the data in different notations often gives access to additional information on the resistance and capacitance of a specific dielectric contribution. In a scenario of a thin film sample, one dielectric contribution may originate from the extrinsic electrode – film interface (Electrode), and one from the intrinsic film (Film). In the case of a parallel-plate capacitor measurement configuration, the two contributions are expected to show spectral features equivalent to two RC elements connected in series. For the different notations the following spectral features are expected as shown from a simulation (Figure 4) for two series RC elements (Figure 4a):

(1.) In plots of $Z''$ vs $Z'$ two semicircles appear, where each semicircle diameter corresponds to the respective RC element resistance $R$, and the angular frequency $\omega_{\max}$ where the semicircle has its maximum is given by $\omega_{\max} = 1/(R{\cdot}C) = 1/\tau$ (Figure 4b).

(2.) $Z''$ vs frequency ($f$) plots show two dielectric relaxation peaks with peak heights of $R/2$ for each RC element resistance (Figure 4c) and the peak frequency is again $\omega_{\max} = 1/(R{\cdot}C) = 1/\tau$. It should be noted that in the literature a dielectric contribution as represented by one RC element is often termed alternatively a "dielectric relaxation".

(3.) $C'$ vs $f$ displays two approximately $f$ independent $C'$ plateaus with values of $C_{\mathrm{Electrode}}$* and $C_{\mathrm{Film}}$*, that correspond approximately to the interface and film capacitance, and shows an intermediate sharp drop by increasing $f$ above a distinct value (Figure 4d).

(4.) $M''$ vs $f$ displays two dielectric relaxation peaks with the peak heights being inversely proportional to each RC element capacitance, $1/C$ (Figure 4e).

(5.) In a Cole-Cole plot of $C''$ vs $C'$ or $\varepsilon''$ vs $\varepsilon'$, one semicircles appears, where the semicircle diameter corresponds to the larger electrode capacitance $C$, and the high-$f$ non-zero intercept with the real axis ($C'$) corresponds to the smaller film capacitance. The angular frequency $\omega_{\max}$ where the semicircle has its maximum is a complicated expression and has little physical significance. Its temperature dependence is sometimes used to reveal dynamic aspects of the phase under investigation (Figure 4f).[6, 25-27]

(6.) $\sigma'$ vs $f$ displays two approximately $f$ independent $\sigma'$ plateaus with values of the approximate electrode and film conductivity, and shows an intermediate sharp increase by increasing $f$ above a distinct value (Figure 4g). In the insets of Figure 4f & 4g it can be noticed that the conversions for $C''$ and $\sigma'$ are quite similar. This implies that $C''$ mainly describes the conductivity of the sample, and it is therefore often referred to as the dielectric losses. $\sigma'$ was calculated here for $g$ = 1 cm, in which case $\sigma'$ can also be denoted as the conductance Y'.

The resistance and capacitance values used for the simulations are given in the Figure 4 caption, where the values were chosen to represent a realistic scenario of one extrinsic electrode interface and one intrinsic film-type contribution. The simulations were carried out using ideal RC elements, whereas real-life IS data is rarely ideal and the capacitors may be replaced by CPEs as mentioned above. If CPEs instead of capacitors are present in the equivalent circuits, this would lead to a suppression of the $Z''$ vs $Z'$ and $C''$ vs $C'$ semicircle centres below the real axis, a broadening of the $Z''$ vs $f$ and $M''$ vs $f$ relaxation peaks and a slight frequency dependence of the $C'$ vs $f$ and $\sigma'$ vs $f$ "plateaus" with a small and constant slope on double-logarithmic axes.[28]

In the framework of the brick work layer model of two RC elements connected in series, the electrode and film capacitance plateaus $C_{\mathrm{Electrode}}$* and $C_{\mathrm{Film}}$* depicted in the $C'$ vs $f$ plot in Figure 4d are both only an approximation for the GB and bulk capacitance $C_{\mathrm{Electrode}}$ and $C_{\mathrm{Film}}$ respectively. In fact, both plateaus are composite terms, where $C_{\mathrm{Electrode}}$* contains parameters from all resistors and capacitors, and $C_{\mathrm{Film}}$* contains a contribution from the electrode capacitance. This is apparent from Figure 4, since $C_{\mathrm{Electrode}}$* and $C_{\mathrm{Film}}$* do not coincide exactly with the values of the capacitors $C_{\mathrm{Electrode}}$ (1 nF) and $C_{\mathrm{Film}}$ (0.1 nF).

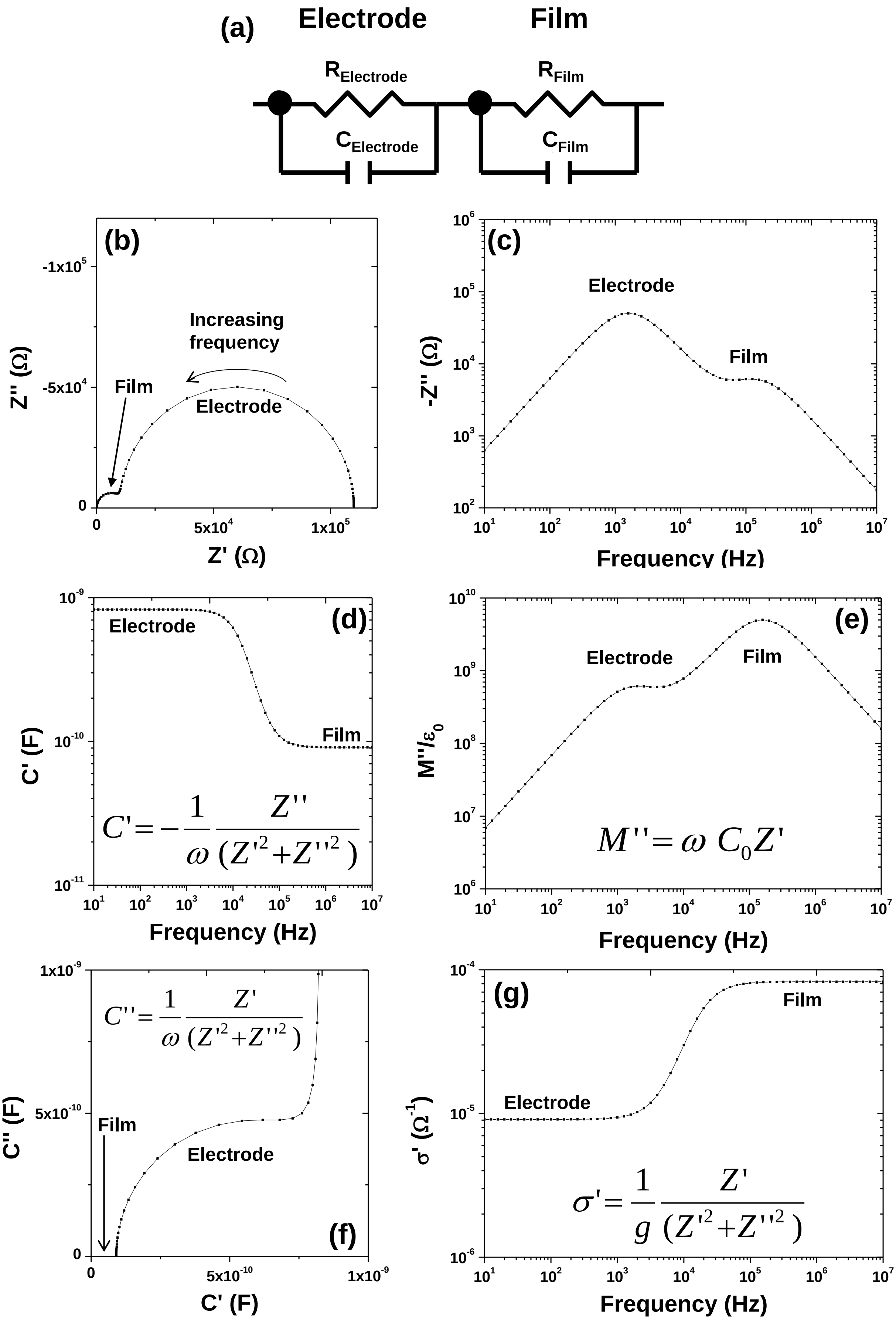

**Figure 4. (a)** Simulation of two series RC elements, presented in different formats: **(b)** $Z''$ (Ω) vs $Z'$ (Ω), **(c)** $Z''$ (Ω) vs frequency ($f$), **(d)** $C'$ (Farad) vs $f$, and **(e)** $M''$ vs $f$. **(f)** $C''$ (Farad) vs $C'$ (Farad), **(g)** $\sigma'$ $(\Omega)^{-1}$ vs $f$. Simulations were carried out with $R_{Electr.}$ = 100 kΩ, $R_{Film}$ = 10 kΩ, $C_{Electr.}$ = 1 nF and $C_{Film}$ = 0.1 nF.

The exact expressions for the plateau values $C_{Electrode}$* and $C_{Film}$* are given by Equations (4) and (5) respectively.

$$C^*_{Electr.} = \frac{R^2_{Electr} C_{Electr} + R^2_{Film} C_{Film}}{(R_{Electr} + R_{Film})^2} \quad (4)$$

$$C^*_{Film} = \frac{C_{Electr.} \times C_{Film}}{C_{Electr.} + C_{Film}}; \quad (5)$$

In the dielectric permittivity notation of $\varepsilon_{Electr.}$*, $\varepsilon_{Film}$*, $\varepsilon_{Electr.}$ and $\varepsilon_{Film}$, Equations (4) and (5) are equivalent. In the case that the electrode and film time constants $\tau_{electr.} = R_{Electr.} \times C_{Electr.}$ and $\tau_{film} = R_{Film} \times C_{Film}$ are sufficiently different by at least 2 orders of magnitude, i.e. $\tau_{electr.} \gg \tau_{film}$, then the plateau values $C_{Electr.}$* and $C_{Film}$* constitute reasonable estimates for the true $C_{Electr.}$ and $C_{Film}$ values. In any other case, especially Equation (4) can have serious consequences. If the capacitance of a sample is measured at a frequency where an extrinsic contribution is dominant, changes in the measured capacitance can be entirely the result of changes in the resistance of the sample. This effect has caused considerable discussion in the field of multiferroic materials, because a measured artificial magneto-capacitance (MC) effect can be entirely the result of magneto-resistance (MR).[21, 29]

The *M*" vs *f* plots shown in Figure 4e are particularly useful to display the intrinsic film relaxation peak. *Z*" vs *f* plots highlight the relaxation peak with the highest resistance *R* (e.g. the electrode peak in Figure 4c), whereas the *M*" vs *f* plots highlight the smallest capacitance (i.e. the film peak in Figure 4e). The approximate peak frequencies for electrode or film peaks as depicted in Figure 4 are given in Equation (6) for the respective plots of *Z*" vs *f* (Electrode) and *M*" vs *f* (Film).

$$f_{max}(-Z'') \approx \frac{1}{2\pi\ R_{Electr.} C_{Electr.}};\quad f_{max}(M'') \approx \frac{1}{2\pi\ R_{Film} C_{Film}} \quad (6)$$

$$-Z''(f_{max}) \approx \frac{R_{Electr.}}{2};\quad M''(f_{max}) \approx \frac{C_0}{2C_{Film}} \quad (7)$$

Equation (6) implies that an extrinsic electrode-type relaxation with large resistance and capacitance would appear at lower *f* than the intrinsic film, which is a common feature in experimental impedance spectra. The different ordinates of the electrode and film peaks are indicated in Equation (7) for a scenario with distinct differences of the time constants between the electrode and film RC elements. The values given in Equations (6) and (7) are approximations, which are getting better with increasingly different time constants of the two dielectric contributions, i.e. $\tau_{electr.} \gg \tau_{film}$. $C_0$ represents the capacitance of the empty measuring cell in vacuum.

The differences in the peak ordinates in Equation (7) guarantee that in *Z*" vs *f* the relaxation peaks representing the largest resistance (Electrode) and in *M*" vs *f* the peaks representing the smallest capacitance (Film) are the most strongly pronounced (see Figure 4 c & e).

In the case where both resistors in Figure 4a are thermally activated and follow Arrhenius type behaviour (e.g. in semiconducting or insulating materials) all the dielectric relaxation peaks, the drop in *C*' vs *f* and the increase in $\sigma$' vs *f* depicted in Figure 4 would all move to higher *f* with increasing temperature, *T*. This appears as if the full spectrum shifts to higher *f* as is demonstrated below in Figure 5b for the example of a $\varepsilon$' vs *f* plot.

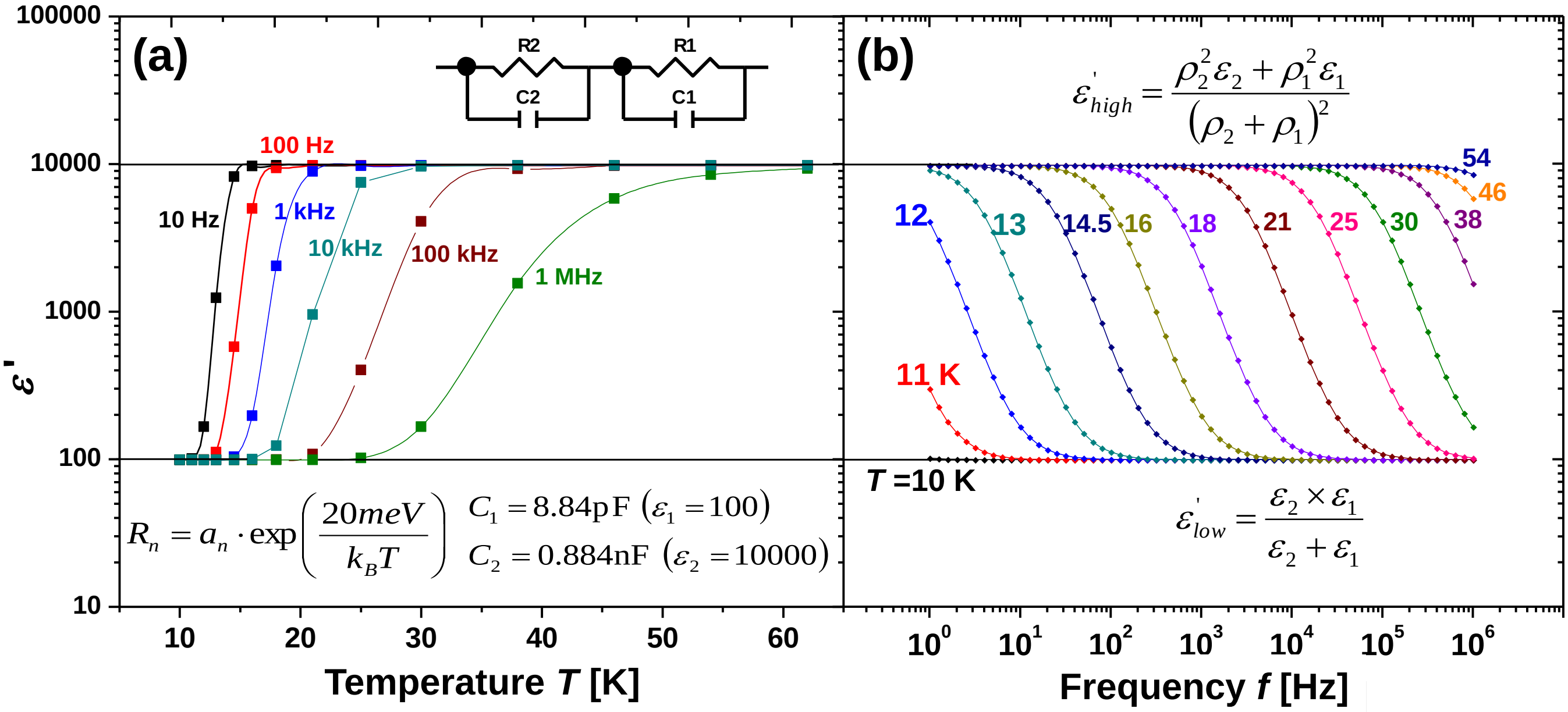

**Figure 5.** Simulated data of **(a)** $\varepsilon$' vs *T* and **(b)** $\varepsilon$' vs *f* data for a series of two RC elements with temperature independent capacitors and Arrhenius activated resistors. Reproduced from Ref. 21 with permission from the American Physical Society (APS).

**f) Temperature dependent IS data**

Figures 5a and 5b show a simulation of the behavior of a series circuit of two RC elements (see Figure 5a inset) in the notations of $\varepsilon'$ vs $T$ (Figure 5a) and $\varepsilon'$ vs $f$ (Figure 5b), where R1-C1 represents the intrinsic film contribution, and R2-C2 the extrinsic electrode interface which is sometimes termed a Maxwell-Wagner contribution.[21] For such $f$ and $T$ dependent simulations the values for the model resistors and capacitors were chosen as indicated in Figure 5a. $T$ independent capacitors C1 and C2, and exponential $T$ dependences of the resistors R1 and R2 were used. The temperature dependence of the resistors was simulated following an activated Arrhenius type behavior, since this is commonly observed in non-metallic materials. Both resistors were assumed to have the same activation energy (20 meV), whereas the pre-exponential terms $a_n$ (see inset Figure 5a) were chosen differently ($a_2 \gg a_1$) such that R2 » R1: $a_1 = 1\ \Omega$, $a_2 = 100\ \Omega$.

The $\varepsilon'$ vs $T$ curves in Figure 5a were calculated from the analytical expression $\varepsilon'(f,R_1,R_2,C_1,C_2)$ for a series of two ideal RC elements for several selected frequencies. The $T$ dependent resistance ($R_n = R_1$, $R_2$) and $T$ independent capacitance ($C_1$, $C_2$) values were substituted into $\varepsilon'(f,R_1,R_2,C_1,C_2)$ using the expressions shown in the inset of Figure 5a. The $\varepsilon'$ vs $f$ curves (Figure 5b) were simply calculated from the expression $\varepsilon'(f,R_1,R_2,C_1,C_2)$ using fixed resistance and capacitance values at each selected temperature (the temperatures in Kelvin are indicated at each curve). $\varepsilon_2$ and $\varepsilon_1$ values represent the extrinsic and intrinsic dielectric permittivity as a representation of the capacitors C2 and C1 respectively, whereas $\varepsilon'$ represents the measured $f$-dependent permittivity. $\rho_2$ and $\rho_1$ values represent the extrinsic and intrinsic resistivity as a representation of the resistors R2 and R1 respectively. Figures 5a & b show two capacitance plateaus, which are consistent with the behavior shown in Figure 4d ($C'$ vs $f$). The time constants $\tau$ of the two RC elements are sufficiently different here (by a factor of $10^4$), such that the two plateau values constitute good approximations for C1 ($\varepsilon_1$) and C2 ($\varepsilon_2$). The exact permittivity plateau values are given in the dielectric permittivity format (Figure 5b inset), which are equivalent to Equations (4) and (5).

Figure 5a clearly demonstrates that a step-like increase of $\varepsilon'$ with $T$ is consistent with the presence of a second extrinsic electrode contribution. The presence of two dielectric contributions and the exponential $T$-dependence of the resistivity in the intrinsic contribution ($R_1$ vs $T$) are sufficient to produce this apparent step-increase. It is important to note that different $R_1$-$T$ dependencies would lead to different features in the $\varepsilon'$ vs $T$ curves. Not only step-like features but basically any variation in $\varepsilon'$ vs $T$ can be the result of a particular intrinsic $R_1$-$T$ relationship.

## III. IS measurement setups for nanomaterials

**a) Parallel-plate capacitor measurement configuration**

Generally, the parallel-plate capacitor measurement configuration is favourable to any other geometry. It allows a clear definition of the electrode distance $d$ and the effective current cross section $A$. Figure 6 shows an example of a semiconducting or insulating thin film, deposited on a conducting substrate and two electrodes are deposited on the film surface. The measurement set-up depicted in Figure 6a implies that the impedance of the film is measured twice across the film along the film normal axis in the out-of-plane direction.[17] The ac current path, illustrated in pink colour, is symmetric in the film areas. On the other hand, the current path in the substrate is rather ill-defined, but this is less important if the substrate is highly conducting and is therefore entirely on the (approximately) “same” electric potential level, i.e. the applied voltage predominantly drops across the film. The electric field lines, represented by black arrows, are approximately parallel.

The fact that they are only approximately parallel is indicated by a small angle of the electric field lines on the edges of the ac current paths indicated, but in thin film samples this effect may be negligible. Therefore, the approximately parallel orientation of the electric field lines allows easy calculation of the specific parameters of resistivity $\rho$ and dielectric permittivity $\varepsilon_r$, where the electrode area corresponds approximately to the effective current cross section $A$ and the electrode distance $d$ is twice the film thickness (the impedance is measured twice across the film). The electrode interface and film contributions are connected in series, and the brick-work layer model would be valid in the case of a homogeneous film. This setup is suitable and commonly used for all types of insulating and semiconducting films. The setup may be inadequate though for more conducting films, where the nominal film resistance would be small in the out-of-plane direction and could possibly not be resolved (the lower resolution limit of IS is typically in the range of 100 Ω, depending on the analyser model). The preferable parallel-plate capacitor measurement setup is often difficult to achieve for nanostructured materials other than thin films. Single molecules, nanotubes or nano-rods on a template or substrate may be difficult to be contacted in an out-of-plane configuration. For thin film structures the measurement setup depicted in Figure 6a is ideal though and may also be used with certain variations:

1.) The conducting substrate may be replaced by an insulating substrate with a pre-deposited metallic electrode layer and the film is deposited on top of the metallic bottom electrode layer.

2.) Instead of using top-top electrodes, it is also possibly to directly contact the conducting substrate or the bottom electrode layer. In this case the impedance would be measured only once across the film in the out-of-plane direction. This is most commonly done by only covering a part of the substrate or the bottom electrode with the thin film material. Part of the substrate or bottom electrode would be exposed to air and can then be contacted easily. Alternatively, a part of the deposited film may be removed by chemical etching or mechanical scratching to expose the substrate or bottom electrode.

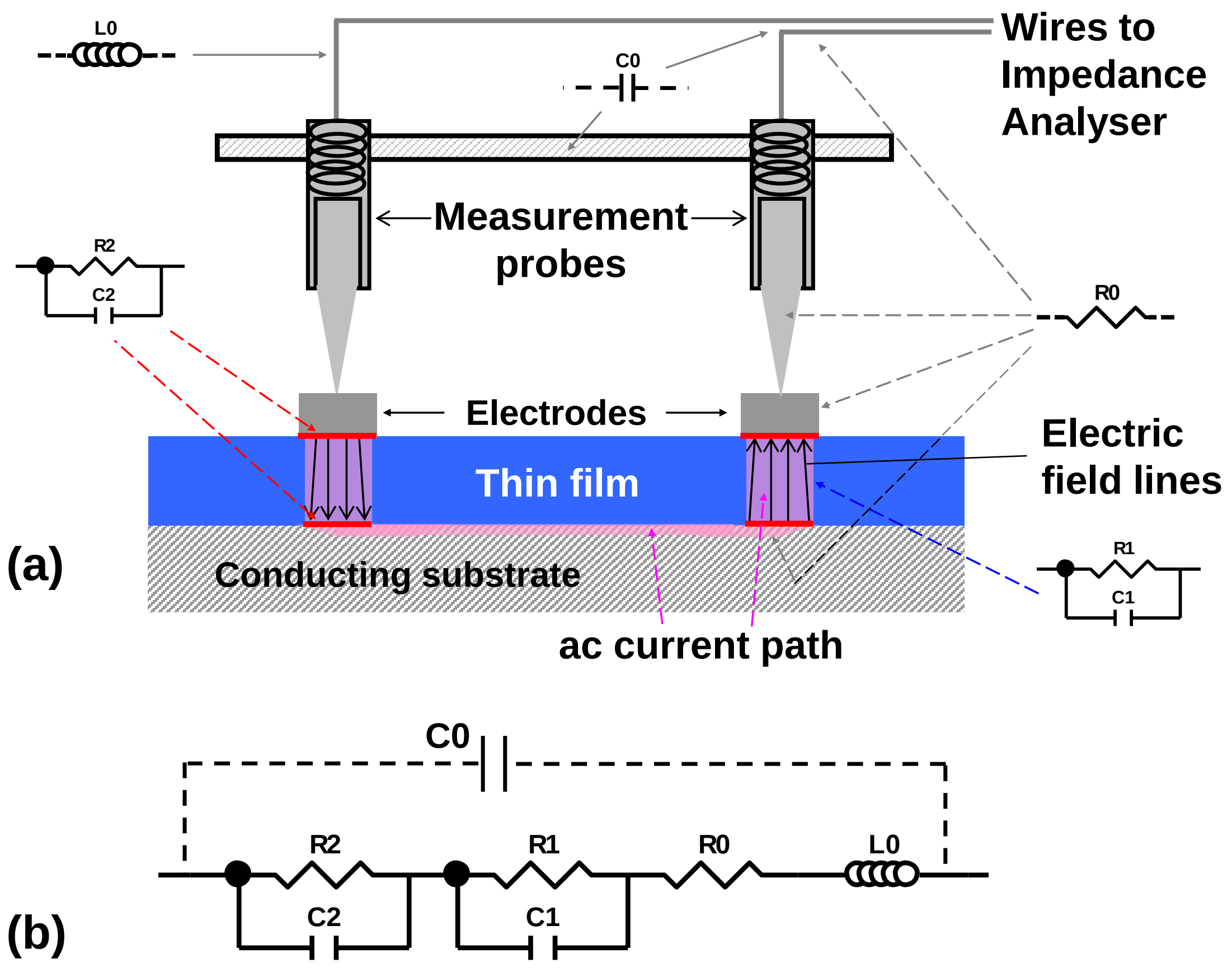

**Figure 6. (a)** Sample geometry for IS parallel-plate geometry measurements of a semiconducting or insulating thin film deposited on a conducting substrate with top-top electrodes **(b)** Equivalent circuit model.

Both alternatives (a) and (b) do not alter the general measurement principle. The equivalent circuit depicted in Figure 6b may be generally valid for parallel –plate capacitor geometries to account for the extrinsic electrode – film interface (R2-C2) and the intrinsic film contribution (R1-C1). The origins of the parasitic contributions C0, R0 and L0 are indicated in Figure 6a, where C0 corresponds to the internal capacitance of the measurement setup (i.e. the probe holder and the cables), whereas R0 and L0 describe the substrate, cable, electrode and probe resistance and inductance respectively. The parasitic capacitance C0 may not always be relevant, because the measurement setup shown in Figure 6a usually leads to large sample capacitance *C* according to Equation (1): *A* has usually medium or macroscopic dimensions, whereas *d* is microscopic. The measurement probes depicted in Figure 6a are equipped with internal springs, which improve the mechanical contact between probes and electrodes.[30]

**b) In-plane measurements**

The measurement and interpretation of IS data is much more troublesome and less well-established in the case of an in-plane direction measurement of a (semi-) conducting film sample, or any other nanostructure that is placed on top of an insulating substrate in between two electrodes. Figure 7a shows an example of a semiconducting or conducting thin film, deposited on an insulating substrate and two top-top coplanar electrodes are deposited on the film surface. In this scenario the electric field lines are not perpendicular to the electrode areas across the entire film. The film and the substrate are in parallel to each other in zone 2, marked in Figure 7a. In zones 1 and 3 the situation is less clear and the film and substrate appear to be in series and parallel at the same time. Since the 2 electrodes are on different electric potential levels during measurements, the electric potential drops between the two electrodes as indicated by the electric field lines. In zone 2, the substrate and the film are both subjected to approximately the same voltage drop due to their parallel alignment.[31, 32] The insulating substrate simply occurs as an additional circuit component in parallel to the film with massively increased resistance. Note that the electric field lines are not affected by the difference in resistance between substrate and film. Charge transport would occur predominantly in the more conducting film, but the substrate would nonetheless contribute to the measured capacitance or even dominate it. The equivalent circuit for zone 2 in the sample consists of two RC elements connected in parallel, as illustrated on the left side of Figure 7b, one for the film and one for the substrate. However, two RC elements in parallel cannot be resolved by IS and the dielectric behaviour of zone 2 is described by only 1 RC element, where the resistor is dominated by the smaller resistance of the two components, i.e. charge carriers move through the more conducting film.

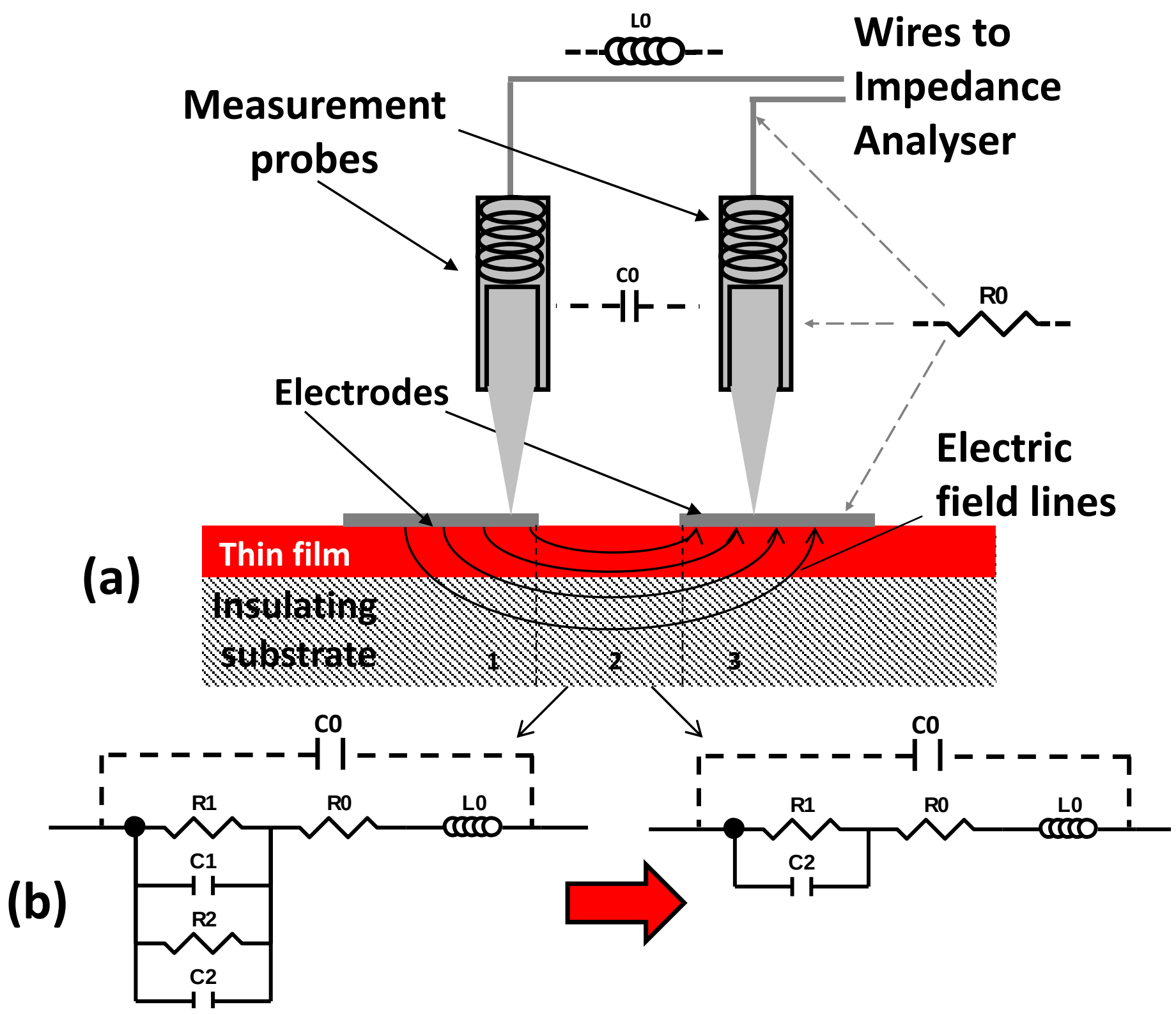

**Figure 7. (a)** Sample geometry for IS in-plane geometry measurements of a (semi-) conducting thin film deposited on an insulating substrate with top-top electrodes. **(b)** Equivalent circuit model of zone 2 in the sample.

On the other hand, the RC capacitance is dominated by the larger capacitance, i.e. the electric dipoles in the substrate polarise and often dominate the measured capacitance due to the macroscopic dimensions of the substrate. It should be noted that the parasitic contribution C0 may also contribute significantly in this measurement setup. Therefore, the dielectric response of zone 2 appears as one dielectric contribution as represented by the RC element in Figure 7b (right side), where the resistor R1 describes charge transport in the more conducting film and the capacitance C2 mainly reflects the polarisability of the substrate and may contain contributions from the parasitic capacitance C0. Figure 7(a) illustrates that the parasitic contributions C0, R0 and L0 appear in the same way as for the parallel-plate capacitor arrangement (Figure 6) and they occur independently to the measurement setups. For the in-plane measurement configuration the parasitic capacitance C0 is more important though, because the sample capacitance is usually low and C0 can therefore contribute significantly. As a general rule, capacitance values measured in the out-of-plane direction are usually larger than capacitance measured with an in-plane measurement setup.
In the case that the two top-electrodes are separated by a macroscopic distance as compared to the film thickness, zone 2 may strongly dominate the impedance spectra. However, there are several indications in the literature that suggest that zones 1 and 3 may also contribute and appear as one additional contribution (in the case that zone 1 and 2 are symmetric), connected in series to zone 2. This additional contribution is expected to show a higher capacitance, because it has a more out-of-plane character. The charge transport in zones 1 and 3 would again be dominated by the more conducting film, but the contributions to the total capacitance from the substrate and parasitic capacitance C0 would be less significant. The equivalent circuit depicted in Figure 7b needed to be extended by one additional RC element in series to R1-C2, i.e. the overall equivalent circuit would be the same as depicted in Figure 6b, but with different associations for the resistors and capacitors. Another third series RC element may be visible in the case that a highly resistive sample electrode interface is present.
This situation appears to be quite complicated, but the temperature dependence of the resistors in zone 1 and zone 2 are expected to be equivalent since they describe the same material (in the case of strictly isotropic charge transport in the in-plane and out-of-plane directions they should be identical).

**c) The geometrical factor**
The dimensions of the measured resistance and capacitance in IS experiments always depend critically on the geometrical factor *g*, which is defined as the effective current cross section *A* divided by electrode distance d . In the case of a parallel plate capacitor measurement setup (Figure 6a), *A* is usually identical to the electrode areas. For an easy calculation of *g* it is therefore preferable to use 2 top electrodes of exactly the same size. For the out-of-plane measurement depicted in Figure 6a, the geometrical factor *g* and concomitantly the measured capacitance *C* are usually large, whereas the resistance *R* is small even for insulating materials.

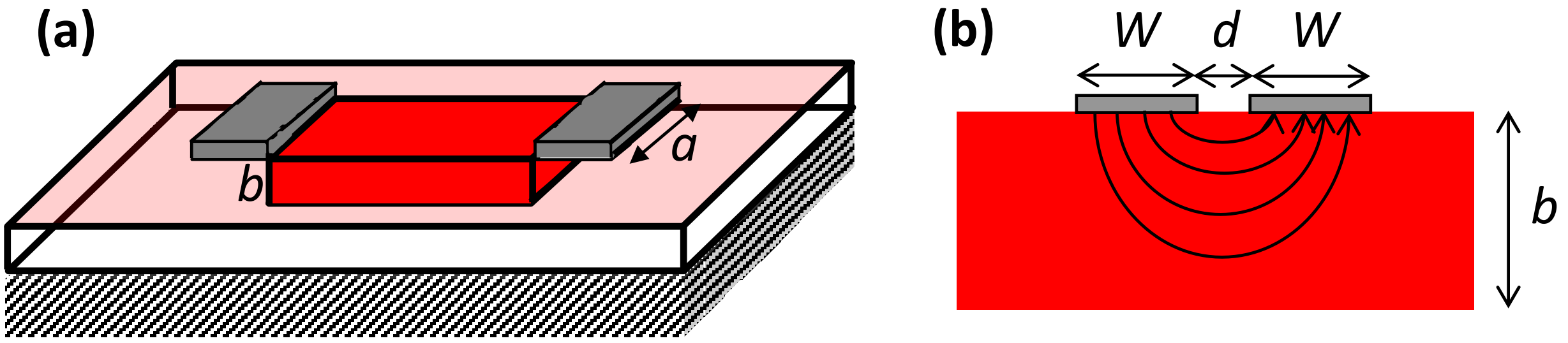

**Figure 8.** Coplanar top-top electrode configurations for in-plane IS measurements.

In the case of an in-plane measurement setup (Figure 7), $g$ is usually small and $\rho$ and $\varepsilon_r$ can only be estimated roughly in the best case. If two coplanar top-top electrodes are separated by a large distance $d$ as compared to the film thickness $b$, $\rho$ may be roughly approximated by using an estimated value for $g = A/d$, where $A$ is given by the film thickness $b$ multiplied by the electrode length $a$, and the electrode distance is $d$. In this approximation the contributions from zones 1 and 3 (Figure 7a) are neglected. This is illustrated in Figure 8a, where the top-top electrodes are shown in dark grey, the film surface area in transparent red, the film interior area is colourless, the substrate is represented by black stripes, and the estimated current path is depicted in dark red. The film permittivity $\varepsilon_r$ may not be accessible in this caseIn the alternative case that $d$ and the electrode width $W$ are small as compared to the film or sample thickness $b$, $\rho$ may not be accessible and the permittivity $\varepsilon_r$ can be estimated as follows:

$$\varepsilon_r \approx \frac{C}{\varepsilon_0} \frac{2K(k)}{K(k')} ; \qquad (8)$$

where $k = d/(2W+d)$, $k' = (1-k^2)^{1/2}$ , and $K$ represents the complete elliptic integral of the first kind. Such an electrode configuration is depicted in Figure 8 (b), which may be difficult to be achieved experimentally. The distance $d$ and width $W$ of the electrodes needed to have micro- or nanoscopic dimensions with a relatively large film or sample thickness $b$ to make Equation (8) a good estimate.

## IV. Experimental IS data from nanomaterials

In the following section several experimental impedance spectra are presented and the respective equivalent circuits are discussed. It is demonstrated in each case that the application of an equivalent circuit to fit impedance spectra allows separately determining the resistance and capacitance from different areas in nanostructured samples. Repeating such equivalent circuit fitting in the same sample at various temperatures and/or applied magnetic fields gives a deep inside into the resistive and capacitive properties of the different areas in the sample. A profound discussion of such dielectric properties goes beyond the scope of this work. The raw experimental impedance spectra in different notations, the equivalent circuit fitting processes and only few extracted parameters are discussed. The subsequent examples have been selected from the author's work.

### a) Out-of-plane magneto-impedance measurements in $BiMnO_3$ thin films

Out-of-plane IS data from a 50 nm $BiMnO_3$ thin film deposited on a conducting 2% Nb-doped $SrTiO_3$ substrate are represented in the $\varepsilon'$ vs $T$ and $\varepsilon'$ vs $f$ notations in Figure 9.[17, 21] The out-of-plane electrode configuration employed was identical to that presented above in Figure 6a using sets of Pt electrodes evaporated onto the layer surfaces. The data resemble the simulations presented in Figure 5 and clear signs of two dielectric relaxation processes are obvious, as manifested by the existence of two $\varepsilon'$ plateaus, $\varepsilon'_{high}$ and $\varepsilon'_{low}$. Each plateau corresponds to one dielectric contribution as represented by one RC element, where the low permittivity plateau $\varepsilon'_{low}$ represents the intrinsic $BiMnO_3$ film permittivity and $\varepsilon'_{high}$ an extrinsic electrode interface relaxation. The data were fitted to the equivalent circuit depicted in the lower inset of Figure 9b. The open symbols in Figure 9 represent the data and solid lines and full symbols correspond to the circuit fit, where excellent agreement between data and model are obvious.

R1-CPE1-C1 corresponds to the $BiMnO_3$ film contribution, whereas R2-CPE2 describes the electrode contribution. Several extrinsic parasitic contributions were accounted for by a series inductor L0 and a resistor R0 describing respectively the inductance and resistance of the measurement wires, conducting substrate and electrodes. L0 and R0 are obvious from the spectra in Figure 9b at the high frequency end, where the curves especially at lower $T$ show a slight upturn. The measured capacitance of the $BiMnO_3$ structure was found to be rather large in the range of nF (nano-Farad), and therefore the parasitic capacitance C0 of approximately 0.2 pF (determined previously by open-circuit measurements) was neglected. The non-ideality of the intrinsic $BiMnO_3$ film relaxations was accounted for by a parallel constant-phase element

(CPE1) and for the extrinsic electrode contribution by the replacement of the ideal capacitor by CPE2.[33]

After rationalizing the presence of extrinsic and intrinsic dielectric relaxations in $BiMnO_3$, the intrinsic magneto-resistance (MR) and magneto-capacitance (MC) can be determined by simply analysing the trends of the intrinsic equivalent circuit components R1 and C1 with magnetic field ($H$), for several fixed temperatures. The complex impedance notation, $Z^* = Z' + i Z''$, is presented in Figure 10 in terms of -$Z''$ vs $Z'$ curves for the $BiMnO_3$ film at 95 K close to its magnetic transition $T_C$, where in such plots a semicircle is expected for each contribution and the semicircle diameter corresponds to the resistivity of the respective contribution as mentioned in section II./e. The partial semicircles in Figure 10a are a manifestation of the intrinsic $BiMnO_3$ film contribution, whereas the extrinsic interface resistance is large and not accessible at this $T$ and $f$ range. Although the intrinsic film semicircle is not fully displayed, the diameter may be extrapolated readily. On increasing $H$ the semicircle diameter decreases, which implies that $\rho_1$ decreases as a result of perceptible MR. The $H$ dependent R1 values obtained from the fittings allowed calculating the intrinsic MR in $BiMnO_3$ films defined as MR = [R($H$=0) - R($H$)]/R(0). The inset panel of Figure 10a shows MR of up to 65% at 95 K and 100 K.

Figure 10b shows $\varepsilon'$ vs $f$ curves for the $BiMnO_3$ film collected under various applied $H$ at 95 K. At low $f$, $\varepsilon'$ shows an upturn, which represents the onset of the extrinsic electrode contribution R2-CPE2 (see $\varepsilon'_{high}$ in Figure 9), where perceptible variations with $H$ occur.

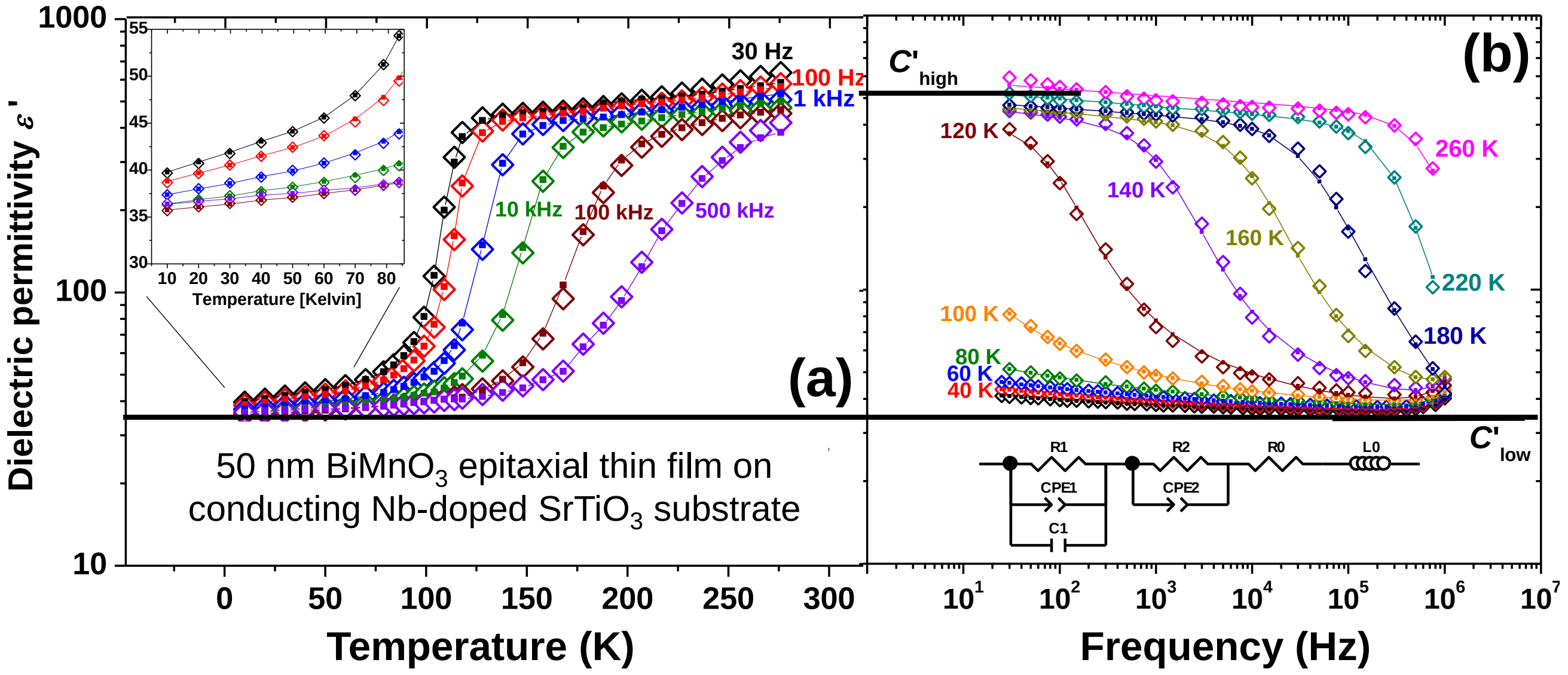

**Figure 9.** Dielectric permittivity **(a)** $\varepsilon'$ vs $T$, and **(b)** $\varepsilon'$ vs $f$ for 50 nm $BiMnO_3$ film at selected frequencies and temperatures as indicated. Open symbols (◊) represent experimental data, full squares (■) and solid lines represent fits to the data using the equivalent circuit depicted in (b). Reproduced from Ref. 21 with permission from the APS

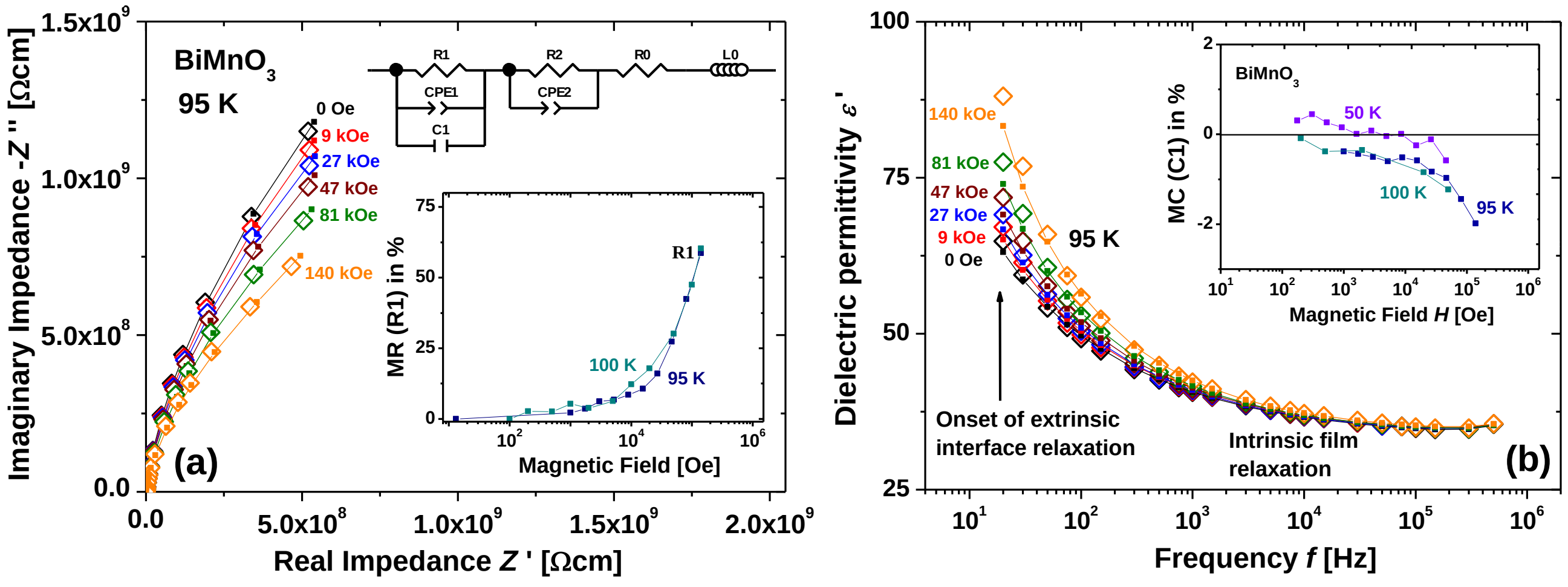

**Figure 10.** $BiMnO_3$ **(a)** imaginary part of the impedance -$Z''$ vs real part $Z'$ and **(b)** $\varepsilon'$ vs $f$, at 95 K for selected applied magnetic fields $H$ as indicated. Open symbols (◊) represent experimental data, full squares (■) and solid lines represent fits to the data using the equivalent circuit depicted in (a). The figure inset panels display **(a)** MR vs $H$, and **(b)** MC vs $H$. Reproduced from Ref. 21 with permission from the APS.

It has been pointed out in section II./e that such variation in $\varepsilon'$ with $H$ in the vicinity of an extrinsic contribution is not necessarily a reflection of MC, but can be caused entirely by the MR of an extrinsic electrode contribution. Since the intrinsic $BiMnO_3$ film relaxation shows perceptible MR, the extrinsic relaxation may be expected to exhibit similar behaviour and the variation in $\varepsilon'$ at low $f$ observed in Figure 10b may well be artificial and simply reflect the MR of the extrinsic electrode contribution. At higher $f$ where the intrinsic relaxation is more dominant, only small MC values can be extracted from the capacitor C1 in the model.

By extracting such C1 values for various $H$, the intrinsic MC, defined as MC = [C($H$=0) - C($H$)]/C(0), can be obtained (inset panel of Figure 10b). The intrinsic MC appears to be rather small in the range of ≈ -1.5% at 90 kOe.[34] In the out-of-plane measurement configuration employed the specific $\rho$ and $\varepsilon_r$ values could be determined readily and all data are therefore represented in normalized specific units.

**b) In-plane IS measurements in $VO_2$ thin films across the metal-insulator transition (MIT)**

In-plane IS data from a 90 nm $VO_2$ thin film grown on an insulating sapphire substrate by sputter deposition are represented in the $\sigma'$ vs $f$ notation in Figure 11.[32] The in-plane electrode configuration employed was identical to that presented above in Figure 7a using pure Indium top-top coplanar electrodes. In this configuration the interpretation of the capacitance is difficult, since the field lines of the applied electric field are not parallel (see Fig. 7a) and $C$ may contain contributions from the film and substrate. Therefore, only the resistive contributions contain unambiguous information on the sample's behaviour, in which case the notation of $\sigma'$ vs $f$ may be advantageous. $VO_2$ undergoes a metal to insulator transition (MIT) upon cooling across the transition $T_C$ ≈ 330 K,[35-37] where microscopic metallic threads appear within the insulating matrix already far below $T_C$.[38]

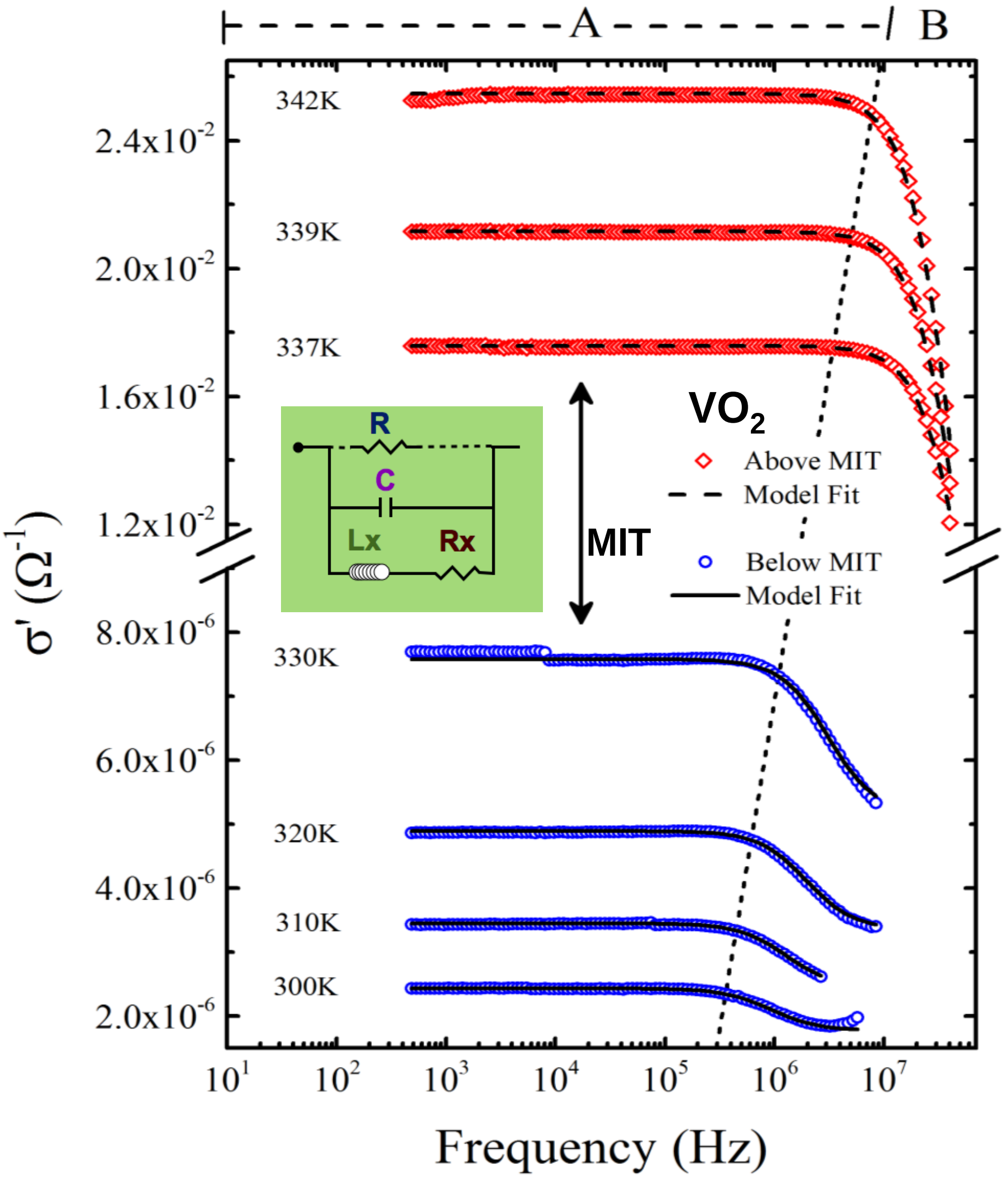

**Figure 11.** IS data collected below and above the MIT, plotted as real part of the conductivity $\sigma'$ vs. $f$. Open symbols indicate the data, solid and dashed lines represent the fit of the data to the model shown in the inset, below and above the MIT. The data fall into two frequency regimes (regimes A and B) as indicated by the dotted line. Reproduced from Ref.32 with permission from AIP Publishing LLC, © 2013.

The macroscopic transition manifests itself by a conductivity jump from $8 \cdot 10^{-6}$ $\Omega^{-1}$ at 330 K to $1.8 \cdot 10^{-2}$ $\Omega^{-1}$ at 337 K as demonstrated in Figure 11 showing data below and above the MIT.In the $\sigma'$ vs $f$ notation, the curves taken above and below the MIT resemble each other, justifying the use of the universal circuit model depicted in the inset of Figure 11.Below and above the MIT the curves show two distinctively different regimes separated by the dotted line in Figure 11: (A) At low/intermediate frequency, the $\sigma'$ vs $f$ curves are approximately $f$ independent, indicative of one standard dielectric contribution as represented by the single RC element in the equivalent circuit (see Figure 4g). The plateau value corresponds to the conductance of the resistor $R$ representing the DC conductance of the sample. (B) At intermediate/high frequency, $\sigma'$ vs $f$ curves show a decrease, which is unusual and inconsistent with a standard RC element. For a second RC element in series a sharp increase in $\sigma'$ vs $f$ would be expected by increasing $f$ (Figure 4g). The behaviour observed here is typical for inductive contributions, leading to decreasing conductivity at high $f$. This inductive behaviour is not consistent though with a serious inductor as expected for a parasitic contribution L0 as described above in section II./d, but can only be modelled with a parallel inductive branch. Therefore, the parallel $L_X$ – $R_X$ branch in the equivalent circuit can be interpreted as a reflection of the inductive behaviour of the metallic threads in the sample, where such threads may be aligned mainly in parallel to the current direction across the insulating areas. Below $T_C$ the highly conductive inductive components are represented by $L_X$ – $R_X$, in parallel to the conventional RC element representing insulating areas. Above $T_C$ the insulating areas no longer contribute to the transport and the resistor R (depicted in dashed lines in Figure 11) can now be eliminated from the circuit. Current now propagates entirely across metallic regions, which are still represented by the $L_X$ – $R_X$ branch. The capacitor C in the circuit is difficult to associate to one specific sample area since C is expected to contain contributions from the film, substrate and parasitic contributions C0. Figure 11 shows that data (coloured symbols) and the equivalent circuit fits (dashed lines above $T_C$ and full lines below) show good agreement. The associations of the equivalent circuit components to certain areas in the sample were all corroborated by the $T$-dependence of such components, i.e the typical $VO_2$ resistance drop for both resistors (R and $R_X$) and a clear drop in $L_X$ across the MIT were confirmed (not shown here).

This example demonstrates that the equivalent circuit modelling process of IS data needs modifications when applied to intrinsically inhomogeneous materials, where the simple RC element model alone cannot describe the data comprehensively. In the in-plane measurement configuration employed here the specific $\rho$ and $\varepsilon_r$ values could not be determined due to the ill-defined effective current cross section $A$ and all data are therefore represented in nominal units.

**c) In-plane IS measurements in $Sm_2CuO_4$ – $LaFeO_3$ multilayers**

Figure 12a shows in-plane IS data in the $Z''$-$Z'$ notation obtained at 260 K from an epitaxial multilayer structure made of 14 unit cells (u.c) $LaFeO_3$ (LFO) and 2 u.c. $Sm_2CuO_4$ (SCO) layers with a repetition rate of 6: $[LFO_{14}/SCO_2]_6$.[39] The 6 repetitions of LFO/SCO were grown on insulating $SrTiO_3$ substrates by sputter deposition, and Ag electrodes were evaporated onto the layer surfaces for IS measurements. The in-plane measurement configuration employed is shown in the Figure 12a inset, where only one repetition of the LFO/SCO layers is depicted. This IS measurement setup is equivalent to that presented above in Figure 7a, but is now employed to a multilayer structure.

LFO and SCO are both insulators, but SCO is well-known to become metallic or even superconducting by replacing $Sm^{3+}$ with $Ce^{4+}$ ($Sm_{2-x}Ce_xCuO_4$), i.e. by electron doping the Cu cations.[40] The intention of fabricating LFO/SCO multilayers was to demonstrate and study the formation of a highly conducting interface between these two nominally insulating oxides by a spontaneous energetically favourable electron transfer from LFO to SCO across the atomically sharp epitaxial interfaces.[41-44]

The three semicircles displayed in Figure 12a were modelled by a series of 3 non-ideal parallel resistor-capacitor (R-CPE) elements (see lower Figure 12a inset). The non-ideal CPE behaviour is evident from the semicircles with their centres being slightly depressed below the $Z'$ x-axis, which indicates a certain distribution of the dielectric time constants $\tau$. Not all three semicircles are fully developed and show considerable overlap, which is highlighted by the dotted model semicircles. Nevertheless, the equivalent circuit model employed (red full symbols and solid lines) shows good agreement with the data (black open symbols). Figure 12b displays the $T$ dependence of all 3 resistor values R1, R2 and R3 obtained from the equivalent circuit fitting procedure, together with the resistance values obtained from two previous measurements of one single phase LFO layer (21 u.c) and one single phase SCO film (5 u.c.). The study of the $T$-dependence of the resistors allows clear association of all 3 contributions to certain areas in the sample: the insulating contribution (R3-CPE3) is associated with an electrode interface and the two conducting contributions (R1-CPE1, R2-CPE2) represent the LFO/SCO interfaces. The resistors R1 and R2 are drastically reduced and display metallic behaviour in the temperature range $T > 120$ K, in contrast to the single phase samples (see Figure 12b). This demonstrates that the SCO near the interface may indeed be electron doped due to a spontaneous electron transfer and becomes conducting. The presence of a large electrode resistance R3 implies that the $T$-dependence of charge transport for the resistors R1 and R2 would not be accessible with DC charge transport measurements, which in fact has motivated the application of IS here in this case to study LFO/SCO interfaces.

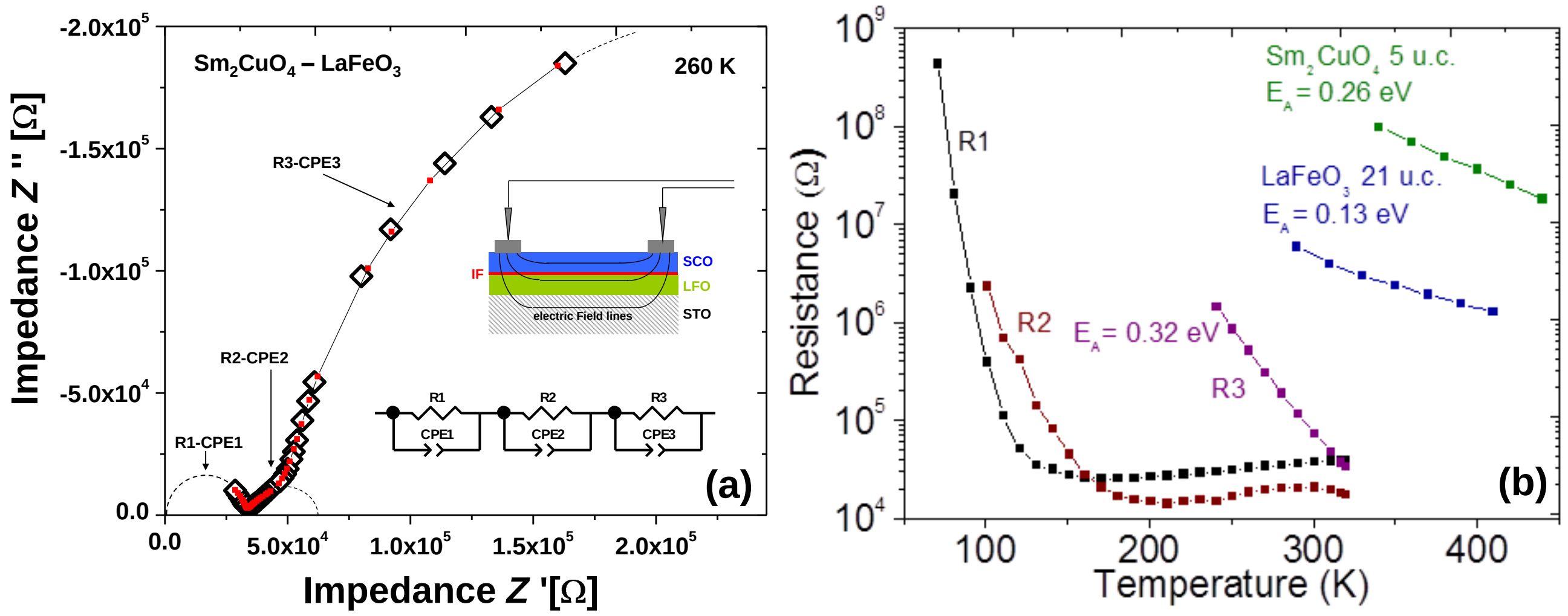

**Figure 12. (a)** IS data displayed as $Z''$ vs $Z'$ for a $[LFO_{14}/SCO_2]_6$ multilayer structure displayed schematically in the upper inset at $T$ = 260K. Open symbols represent data points, red solid squares and thin solid lines represent fits using the model shown in the lower inset. **(b)** Temperature dependence of the resistance values obtained from the equivalent circuit resistors R1, R2 and R3. Reproduced from Ref. 39 with permission from WILEY-VCH Verlag, © 2013.

The association of the 3 semicircles to certain areas in the multilayer structure was supported by the capacitance values obtained from the fits shown in Figure 13. The capacitance C1 is rather low in the range of 2 – 20 pF with a clear temperature dependence, reminiscent of the $SrTiO_3$ behaviour. Therefore, C1 may be associated with a contribution from the substrate, whereas the resistor R1 originates from the electron doped LFO/SCO interface.

Such a scenario is typical for an in-plane measurement setup as mentioned above in section III./b, where the conducting film contribution is connected in parallel to the substrate, and one single R1-CPE1 element appears in the spectra with R1 being dominated by the conducting interface resistance and CPE1 (C1) by the $SrTiO_3$ substrate capacitance. In the case of $SrTiO_3$ with a relatively large dielectric permittivity it is no surprise that C1 is dominated by the substrate and parasitic contributions C0 or thin film contributions are not dominant. The element R1-CPE1 may therefore be associated with a zone 2-type dielectric contribution as indicated in Figure 7a.

The capacitance C2 is significantly increased (3 – 5 nF) as compared to C1 without a clear temperature dependence. Therefore, C2 may be dominated by LFO and SCO areas in the film, possibly near the electrodes. This is corroborated by the resistor R2, which still originates from the electron doped LFO/SCO interface and the element R2-CPE2 may therefore be associated with a zone 1/3-type dielectric contribution as indicated in Figure 7a.

The capacitance C3 is further increased (30 – 40 nF) as compared to C2, again without a clear temperature dependence but the resistor R3 is now insulating. This and the fact that C3 is the largest overall capacitance leads to the conclusion that the element R3-CPE3 should be associated with an electrode sample interface contribution.

The metallic behaviour of R1 and R2 above 120 K and the sharp resistance upturn below may be ascribed to a metal insulator transition occurring in the doped SCO layer near the interface. This behaviour is typically found in underdoped non-superconducting cuprates,[45] where the doping level may be insufficient to make the sample superconducting. In the similar $Nd_{2-x}Ce_xCuO_4$ system the phase diagram suggest that a doping level of $0.05 < x < 0.14$ in the SCO near the interface may be likely.[45]

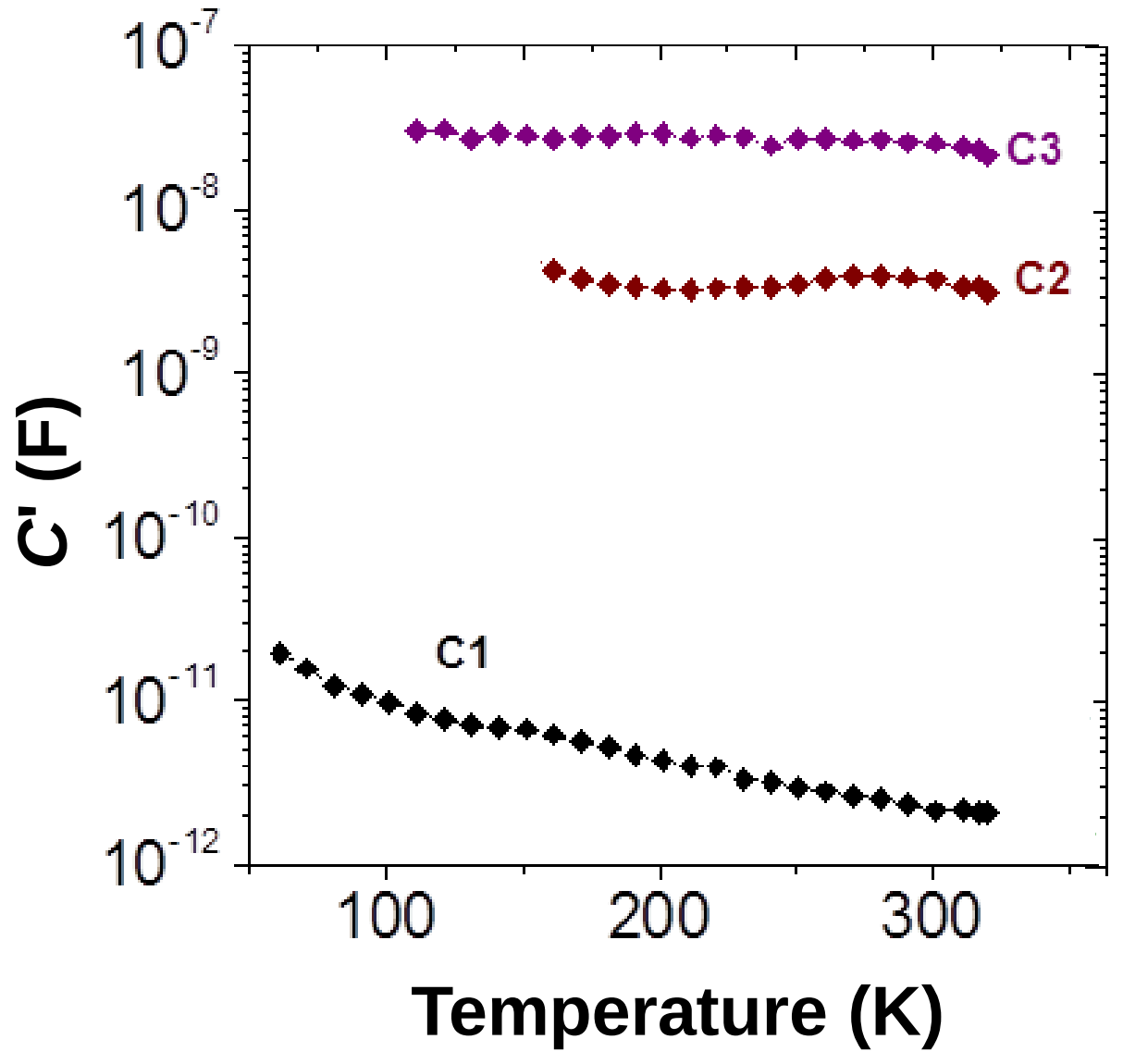

**Figure 13.** Temperature dependence of the capacitance values obtained from the equivalent circuit capacitors C1 (CPE1), C2 (CPE2) and C3 (CPE3). Reproduced from Ref. 39 with permission from WILEY-VCH Verlag, © 2013.

**d) Out-of-plane IS measurements in $LaMnO_3$ – $SrTiO_3$ multilayers**

Figure 14 shows out-of-plane IS data in the *M*'' vs *f* and *C*' vs *f* notations obtained at 140 K from an epitaxial multilayer structure made of 15 u.c $LaMnO_3$ (LMO) and 2 u.c. $SrTiO_3$ (STO) layers with a repetition rate of 8: $[LMO_{15}/STO_2]_8$.[46-48] The 8 repetitions of LMO/STO were grown on a conducting 2% Nb doped $SrTiO_3$ substrate by sputter deposition. Ag electrodes were evaporated onto the layer surface and Al/Ag electrodes on the bottom of the Nb:STO substrate for out-of-plane IS measurements. The out-of-plane electrode configuration employed is shown in Figure 14a, where only 4 repetitions of the LMO/STO layers are depicted. This IS measurement setup is similar to that presented above in Figure 6a, but is now employed to a multilayer structure.

In the multilayer structure presented the capacitance values detected are all rather small (see Figure 14c inset) and cannot be explained by a strict out-of-plane alternating current (AC) direction. Since the LMO layers are expected to be semiconducting and the STO would be insulating, AC charge transport may not occur strictly in the out-of-plane direction. The AC currents may flow predominantly in the in-plane direction along the semiconducting LMO layers, but must pass across the insulating STO layers in the out-of-plane direction in some way. This crossing occurs most likely through pin-holes in the rather thin (2 u.c.) STO layers and the AC currents may meander within the LMO planes from one STO pin-hole in the first STO layer to the next pin-hole in the consecutive STO layer. This rather complex situation is illustrated in Figure 14a and a non-standard equivalent circuit is required as depicted in Figure 14b. Figure 14 c demonstrates that this equivalent circuit fitting leads to good agreement of the data (open symbols) with the model (full symbols and solid lines).

In the equivalent circuit model depicted in Figure 14b, the $R_{IF}$-$C_{IF}$ contribution is interpreted as a large charge blocking Schottky barrier forming between the first LMO and the Nb:STO substrate, because the resistor $R_{IF}$ had to be set to infinity for reasonable fitting of the data obtained within the *f*-range available, i.e. total charge blocking was indicated. This charge blocking Schottky type barrier is in fact well-known to occur between manganites and Nb:STO.[49] For an in-plane (ip) electrical current flow within the LMO layers a competition between the parallel LMO and STO layers is expected, and the impedance spectra are dominated by the lowest resistance (in this case LMO: $R_{ip}$), i.e. the current flows within the LMO layers, and the highest capacitance (in this case STO: $C_{ip}$). It should be noted that the LMO and STO contributions can still be modelled by one RC element each, but the LMO capacitance and STO resistance (both depicted by dashed lines in Figure 14b) are not relevant and can be eliminated from the circuit as mentioned above in section III./b (see also Figure 7b) due to the parallel alignment of LMO and STO layers. The areas in the sample where the AC currents pass across the insulating STO layers through pin-holes in the out-of-plane (oop) direction are described by the $R_{oop}$-$C_{oop}$ element. Figure 14b only depicts the Nb:STO substrate and the first LMO/STO bilayer repetition and the equivalent circuit in this first bilayer area is assumed to be representative of all 8 repetitions.

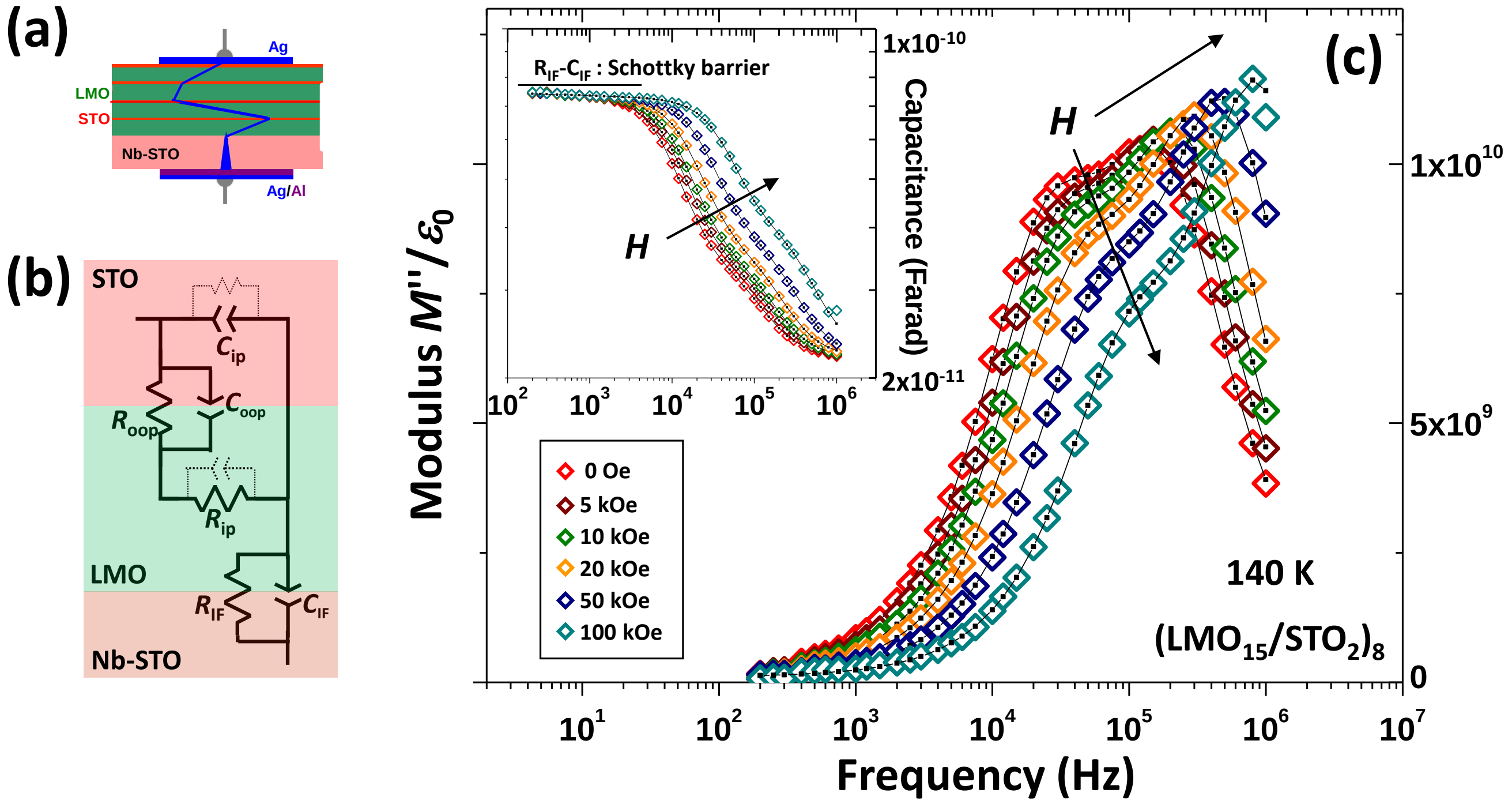

**Figure 14. (a)** Meandering-type charge transport across $[LMO_{15}/STO_2]_8$ multilayers. **(b)** Equivalent circuit model containing in-plane ($R_{ip}$-$C_{ip}$) and out-of-plane ($R_{oop}$-$C_{oop}$) contributions, and a blocking type interface barrier ($R_{IF}$-$C_{IF}$). **(c)** *f* and *H* dependent *M*'' vs *f* dielectric data. Open symbols (◊) correspond to the data, solid lines and squares (▪) represent fits to the model in (b).

In fact, all capacitors in the circuit were represented by CPEs, and the CPE capacitance values extracted from the fits were converted to real capacitance by the standard procedure mentioned above. The associations of the equivalent circuit components to certain areas in the sample were all corroborated by the *T*-dependences of such components, i.e. typical LMO, STO or interfacial behavior (not shown here).

Figure 14c depicts the IS data in the *M*'' vs *f* and *C*' vs *f* notations obtained at 140 K under various applied magnetic fields (*H*) as indicated in the Figure 14c lower inset. The advantage of the *M*'' vs *f* notation is clearly demonstrated here, because the large $C_{IF}$ capacitance is not visible. This is because the ordinates of the dielectric relaxation peaks displayed in *M*'' vs *f* are inversely proportional to the capacitance of the respective contribution as mentioned above in section II./e. Therefore, the intrinsic $R_{ip}$-$C_{ip}$ and $R_{oop}$-$C_{oop}$ contributions can be analysed much easier in the *M*'' vs *f* notation as compared to the *C*' vs *f* plots (Figure 14c inset), where in the latter the large $C_{IF}$ capacitance dominates. The double peak structure shown in *M*'' vs *f* shows significant variations of the peak heights with applied *H*, which corresponds to clear MC effects.

Due to the complex effective current cross section *A* and an LMO/STO bilayer repetition rate of 8 no specific units could be estimated.

## V. Conclusions

The dielectric response of solid matter can generally be represented by standard RC element components for each dielectrically distinct area in the sample. Whereas such RC elements are connected in series for macroscopic bulk materials, nanostructured materials are often arranged in a way such that the strictly series connection no longer applies. In this case, more complex equivalent circuits have to be developed, where the association of the equivalent circuit components to certain areas in the sample can be done most effectively by considering the sample architecture and the resulting trends of the circuit components with temperature and/or magnetic field. Impedance spectroscopy is undoubtedly the method of choice to disentangle multiple contributions to the resistive and capacitive properties of complex nanostructured materials.

## Acknowledgements

The author wishes to acknowledge the invaluable help with sample preparation, impedance spectroscopy measurement setup, software programming, data analysis and useful discussions provided by his colleagues Eric Langenberg, Jofre Ventura, Javier Garcia-Barriocanal, Flavio Y. Bruno, Gabriel Ramirez, Fabian A. Cuellar, Norbert M. Nemes, Neven Biskup, Maria Varela, Alberto Rivera-Calzada, Zouhair Sefrioui, Carmen Munuera, Edgar J. Patiño, Carlos Leon, Manuel Varela, Mar Garcia-Hernandez, Ivan K. Schuller and Jacobo Santamaria.

The MICINN/MINECO in Spain is acknowledged for granting a Ramón y Cajal fellowship.